\documentclass[9pt,reprint, amsmath,amssymb, aps,
pra,
]{revtex4-2}

\newcommand{\tr}{\textup{tr}}
\newcommand{\0}{{(0)}}
\newcommand{\1}{{(1)}}
\newcommand{\2}{{(2)}}

\usepackage{graphicx}
\usepackage{dcolumn}
\usepackage{bm}

\usepackage{amsmath,amsfonts,amssymb}
\usepackage{graphicx,textcomp}
\usepackage{algorithm,algpseudocode}
\usepackage[colorlinks=true, allcolors=blue]{hyperref}
\usepackage{multirow}
\usepackage{arydshln}
\usepackage{caption}

\usepackage{soul}
\usepackage{xcolor}

\newcommand{\vnu}{{\vec{\nu}}}

\newcommand{\betab}{{\boldsymbol{\beta}}}
\newcommand{\bsigma}{{\boldsymbol{\sigma}}}

\newcommand{\cbeta}{\beta}
\newcommand{\bij}{\lambda}

\begin{document}

\preprint{APS/123-QED}

\title{Protocol for suppression of noise from stimulated multi-photon emissions in concatenated entanglement swapping links and quantum repeaters}

\author{Yousef K. Chahine}
\email{yousef.k.chahine@nasa.gov}
\author{Ian R. Nemitz}%
\author{John D. Lekki}%
\affiliation{NASA Glenn Research Center, Cleveland, OH}

\begin{abstract}
Multi-photon emissions constitute a fundamental source of noise in quantum repeaters and other quantum communication protocols when probabilistic photon sources are employed.   In this paper, it is shown that by alternating the Bell state measurement (BSM) basis in concatenated entanglement swapping links one can automatically identify and discard many errors from stimulated multi-photon emissions.  The proposed protocol is shown to completely eliminate the dominant quadratic growth of multi-photon errors with the length of the repeater chain.  Furthermore, it is shown that the protocol can be employed in satellite-assisted entanglement distribution links to enable links which are more robust in the presence of imbalanced channel losses.  The analysis introduces a convenient calculus based on Clifford algebra for modeling concatenated entanglement swapping links with multi-photon emissions.  In particular, we present a compact expression for the fidelity of the Bell state produced by a repeater chain of arbitrary length including noise from double-pair emissions.
\end{abstract}

\maketitle

\section{Introduction}

The ability to establish shared quantum systems exhibiting entanglement between remote locations has a growing number of potential applications, including quantum clock synchronization \cite{KOMAR2014,NICHOL2022}, distributed quantum sensing \cite{ZHANG2021,GOTTESMAN2012}, and experimental tests of fundamental physics in new regimes \cite{MOHAGEG2022}.  The quantum repeater protocol enables the distribution of entangled states over long distances by concatenating elementary links using entanglement swapping.  In this work, we study repeater protocols with minimal quantum processing including probabilistic BSMs, quantum memories, and multiplexing with active mode switching across spatial, spectral, or temporal modes \cite{SINCLAIR2010,SANGOUARD2011,GUHA2015,PANT2017}.

It was shown in \cite{GUHA2015} that multi-pair emissions can severely limit useful application of parametric down conversion (PDC) sources in such a repeater architecture.  Photon number-resolving (PNR) detectors can be employed to overcome this limitation \cite{KROVI2016}; however, effective PNR requires highly efficient repeaters, since any photon losses between multi-photon production and detection limit the ability of the detector to identify multi-photon emissions.  In this paper, we introduce an alternating Bell state measurement (ABSM) protocol for further mitigating multi-photon noise in concatenated entanglement swapping links employing PDC sources which does not rely on PNR detection.  The protocol exploits correlations in the two-mode squeezed vacuum (TMSV) state to automatically identify stimulated double-pair emissions at adjacent BSMs.  The ABSM protocol thus alleviates the impact of imperfect PNR caused by inefficiencies between the source and linear optical BSM.  In particular, it is shown in Section \ref{sec:repeaters} that the ABSM protocol eliminates the dominant quadratic contribution to the growth of multi-photon errors in extended repeater chains with PDC sources.

The ABSM protocol also mitigates the increase in multi-photon noise in elementary links with imbalanced channel losses (Sections \ref{sec:repeaters}-\ref{sec:absmgain}).  Imbalanced channels result from practical limitations and are unavoidable for satellite-assisted links and dynamic links where the locations of source nodes, BSM nodes, and repeater stations are constrained.  If probabilistic photon sources are employed, the multi-photon emissions from one source can dominate the BSM, requiring attenuation of the source closer to the BSM to reduce the number double-pairs.  This additional attenuation effectively rebalances the channel losses at the cost of a reduction in link efficiency.  In Section \ref{sec:repeaters}, it is shown that the fidelity of the Bell state produced using the ABSM protocol is more robust to imbalanced losses.  Furthermore, the ABSM protocol can be used in conjunction with multiplexed, cascaded PDC sources \cite{DHARA2022} to further suppress double-pairs in imbalanced entanglement swapping links (Section \ref{sec:absmgain}).

The suppression of multi-photon noise is also important for concatenation of elementary links employing a DLCZ-type protocol using atomic ensembles \cite{SANGOUARD2008,SANGOUARD2011,YU2020}.  Sources based on atomic ensembles produce entangled states exhibiting bosonic excitation statistics with a Hamiltonian formally identical to that of PDC \cite{SANGOUARD2011}, and thus can suffer the same types of multi-photon errors when employed in a quantum repeater chain.  However, various DLCZ-type protocols based on two-photon interference include an additional detection entangling the emissions from several atomic ensembles leading to a more complex form for the multi-photon term, the analysis of which is beyond the scope of this work \cite{SANGOUARD2008}.

The ABSM protocol is presented in Section \ref{sec:protocol}, where it is shown that by employing alternating measurement bases in the linear optical BSMs, a stimulated double-pair emission from a single source cannot falsely trigger two adjacent BSMs.  The resulting suppression of multi-photon errors is analyzed in Section \ref{sec:repeaters}, where it is shown that these errors constitute the dominant source of noise in extended quantum repeater chains using probabilistic sources, growing quadratically with the number of elementary links $\ell$ using the standard repeater protocol.  The analysis is based on a new calculus introduced in Section \ref{sec:linkmodel} for key observables of the quantum state produced by repeater chains including multi-photon terms, enabling a closed form expression for the Bell state fidelity produced by a repeater chain of arbitrary length.  This simplifies the analysis compared to other approaches based on the Wigner formalism for Gaussian states \cite{TAKEOKA2015} or numerical density matrix calculations \cite{KHALIQUE2013,GUHA2015}.  Other potential applications of the ABSM protocol for satellite-assisted entanglement distribution are discussed in Section \ref{sec:absmgain}.

\section{The alternating BSM protocol}\label{sec:protocol}

\begin{figure}[h!]
\centering\includegraphics[width=8.7cm]{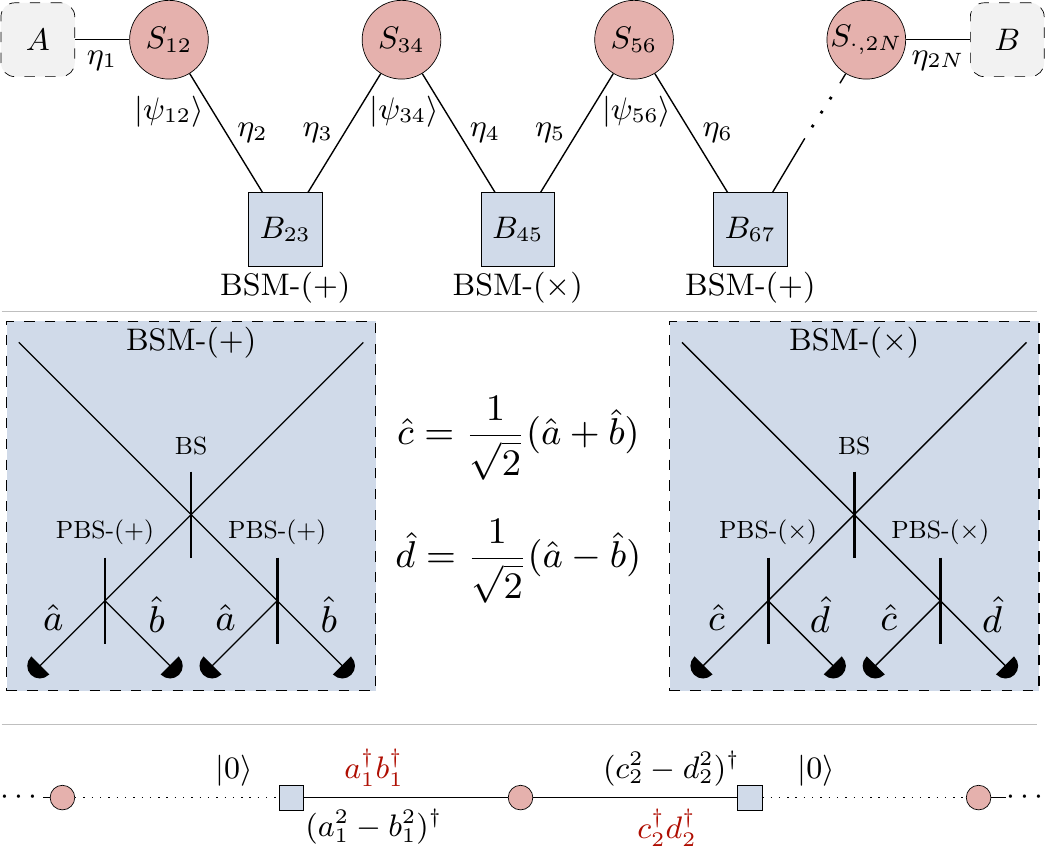}
\captionsetup{font=footnotesize,labelfont=footnotesize,justification=raggedright}
\caption{Protocol for concatenated entanglement swapping links.  Red circles denote entangled pair sources $S_{ij}$ and blue squares denote Bell state analyzers $B_{ij}$.  Channels are labeled by the corresponding transmission efficiency $\eta_i$.  By alternating the BSM measurement basis, a double-pair emission from a single source $S$ cannot trigger two adjacent BSMs.}
\label{fig:protocol}
\end{figure}

The protocol is based on the observation that the 4-photon emission in a polarization-entangled TMSV state exhibits certain correlations between the 2-photon states emitted into each output channel \cite{KOK2000,LAMASLINARES2001}.  Namely, measurement of a 2-photon state with orthogonal polarizations in one output channel is correlated to a 2-photon state with correlated polarizations in the opposite channel when measured in a diagonal basis \eqref{eq:4photoncorr}.  Since a linear optical BSM requires detection of two photons with opposite polarizations, by alternating the measurement basis in a concatenated entanglement swap, a 4-photon emission from a single source cannot independently trigger two adjacent BSMs (Fig. \ref{fig:protocol}).

To make this rigorous, we express the entangled photon state---representing a pair of dual-rail qubits in a superposition of modes with bosonic annihilation operators $\{a_1,b_1\}$ and $\{a_2,b_2\}$---in the form \cite{DURKIN2002,GUHA2015}
\begin{align}\label{eq:state}
\begin{split}
|\psi_{12}\rangle &= \sqrt{p^{(0)}_{12}}|0\rangle + \sqrt{p_{12}^\1 /2}\big(a_1 b_2 - b_1 a_2 \big)^\dagger|0\rangle \\
& +  \sqrt{p^\2_{12}/12}\big(a_1^2 b_2^2 + b_1^2 a_2^2 - 2a_1 b_1 a_2 b_2\big)^\dagger|0\rangle
\end{split}
\end{align}
where we have truncated higher order terms in order to focus on the lowest order contribution to the multi-photon error.  This form for the entangled pair state is a tensor product of two copies of a TMSV state derived from the Hamiltonian for a PDC process.  For the full TMSV state, the probability $p_{12}^{(n)}$ of producing $n$ pairs is given by
\begin{equation}\label{eq:npaird}
p^{(n)}_{12} = (n+1)(1-|\lambda|^2)^2|\lambda|^{2n}
\end{equation}
where the parameter $|\lambda|^2\ll 1$ determines the single-pair emission probability $p_{12}\equiv p_{12}^\1 \simeq 2|\lambda|^2.$ In order to normalize the truncated state without changing the relation between single- and double-pair emissions, we put $p_{12}^\0 = 1-p_{12}^\1 - p_{12}^\2$.  This normalization relies on the double-pair term to represent the multi-photon noise and results in a small overestimation of the vacuum component which subsumes the contribution from higher-order photon emissions \cite{DHARA2022}.  

Correlations in the 4-photon term arise from the stimulated emission process captured in the bosonic relation $a^\dagger|n\rangle = \sqrt{n+1}|n+1\rangle$ which leads to a suppressed mixed term
\begin{equation}\label{eq:4photonstate}
|2,0;0,2\rangle + |0,2;2,0\rangle - |1,1;1,1\rangle
\end{equation}
when written in the basis of Fock states $|m_1,n_1;m_2,n_2\rangle$ in modes $\{a_1,b_1,a_2,b_2\}$.  A less obvious correlation is found by expanding the state of modes $\{ a_2,b_2\}$ in the diagonal basis $c_2 = \sqrt{1/2}(a_2 + b_2)$ and $d_2 = \sqrt{1/2}(a_2 - b_2)$ to obtain the 4-photon term in the form
\begin{align}\label{eq:4photoncorr}
\begin{split}
(a_1^2)^\dagger(c_2^2 + d_2^2 - 2c_2 d_2)^\dagger |0\rangle & \\
+(b_1^2)^\dagger(c_2^2 + d_2^2 + 2c_2 d_2)^\dagger |0\rangle & \\
-2(a_1 b_1)^\dagger(c_2^2 - d_2^2)^\dagger |0 \rangle &.
\end{split}
\end{align}
The key property is that a measurement of opposite polarizations $a_1,b_1$ in the first channel projects the state in the second channel onto a correlated 2-photon N00N state in modes $c_2,d_2$ (and vice-versa).

One can exploit this correlation in a concatenated entanglement swap by noting that a linear optical BSM between two adjacent sources $|\psi_{12}\rangle|\psi_{34}\rangle$ only succeeds if opposite polarizations are detected.  Furthermore, the BSM can be performed in either basis, as follows from the easily verified relations
\begin{align}\label{eq:bsmapping1}
|\Psi^+\rangle_{ab} &= |\Phi^-\rangle_{cd},  &|\Phi^+\rangle_{ab} = |\Phi^+\rangle_{cd}, \\
|\Psi^-\rangle_{ab} &= -|\Psi^-\rangle_{cd}, &|\Phi^-\rangle_{ab} = |\Psi^+\rangle_{cd},\label{eq:bsmapping2}
\end{align}
where
\begin{align}
|\Psi_{ij}^\pm\rangle_{ab} &= \sqrt{1/2}(a_i^\dagger b_j^\dagger \pm b_i^\dagger a_j^\dagger)|0\rangle, \\
|\Phi_{ij}^\pm\rangle_{ab} &= \sqrt{1/2}(a_i^\dagger a_j^\dagger \pm b_i^\dagger b_j^\dagger)|0\rangle 
\end{align}
represent the dual-rail Bell states in the $\{a,b\}$ basis, with analogous expressions for the representation in the $\{c,d\}$ basis.  Thus, by performing adjacent Bell state measurements in concatenated entanglement swapping links in alternating diagonal bases, as illustrated in Figure \ref{fig:protocol}, the correlations observed in \eqref{eq:4photoncorr} ensure that a multi-pair emission cannot independently trigger a BSM in two adjacent Bell state analyzers.

Two important caveats are immediately apparent.  First, the protocol relies on the correlations in the multi-photon state described by \eqref{eq:state}.  Results presented in \cite{LAMASLINARES2001} give experimental justification for this form for the 4-photon term, although source impurities allowing emission of each photon into more than two modes may degrade the correlations in the double-pair emissions.  For example, if the secondary photon pair is emitted into a set of orthogonal modes $\{a_1',b_1',a_2',b_2'\}$ (e.g. an adjacent temporal mode), then the required correlations do not exist. However, this is not a problem in practice since any implementation of the elementary entanglement swapping link must already filter out adjacent-mode noise in order to ensure the single-pair emissions yield indistinguishable photons at the beam-splitter where the BSM is performed.  A more fundamental limitation of the protocol is the fact that it does not suppress multi-photon errors when one of the photons in the double-pair state is lost.  In this case, the chain can still be corrupted if a photon from an adjacent source arrives at the BSM where the secondary emission was lost, although this type of error can still yield a Bell state with 75\% fidelity (\emph{cf.} Eq. \eqref{eq:dpfidelity}).  In Sections \ref{sec:repeaters}-\ref{sec:absmgain} it is shown that the protocol nevertheless provides significant multi-photon error mitigation for extended repeater chains as well as for elementary links with imbalanced losses.

\section{Concatenated link model}\label{sec:linkmodel}

In order to analyze the noise suppression and gain afforded by the ABSM protocol we model a concatenated entanglement swapping link as shown in Figure \ref{fig:protocol}.  We assume completely dephasing, pure-loss bosonic channels $ a_i = e^{i\phi}\sqrt{\eta_i} a_i' + \sqrt{1-\eta_i} a_i''$ with transmission efficiency $\eta_i$, where $\phi$ is a random phase accumulated equally on modes $a_i$ and $b_i$.  For simplicity, we assume that the external channels connecting the sources $S_{12}$ and $S_{2N-1,2N}$ to the receivers at $A$ and $B$ are lossless $\eta_1=\eta_{2N}=1$.  Detection in each linear optical BSM is modeled via the detector POVM including PNR as in \cite{KROVI2016}; however, to simplify the analysis we neglect extrinsic noise (i.e. we assume no background/dark counts).  For all interior channels $\eta_2,...,\eta_{2N-1}$ the detection efficiency can be grouped with the channel transmission efficiency---the justification for this combined efficiency is an equivalence of quantum operations discussed in the Supplementary Material (SM).  The detector POVMs are combined to form a single POVM consisting of the four possible successful outcomes from a chain of nominally successful BSMs---corresponding to the states $|\Psi_{AB}^\pm\rangle$ or $|\Phi_{AB}^\pm\rangle$---together with the complementary outcome representing failure of the link.  Note that if all of the sources produce the state $|\psi_{ij}\rangle$ corresponding to $|\Psi^-_{ij}\rangle_{ab}$, then the non-alternating BSM protocol will only produce the outcome corresponding to $|\Psi_{AB}^\pm\rangle$ at the output of the full concatenated link; however, it follows from \eqref{eq:bsmapping1}-\eqref{eq:bsmapping2} that all four outcomes are possible when employing the ABSM protocol.

We now develop the expression for the link efficiency and Bell state fidelity for a passively concatenated link---at the end of the section we discuss how these results directly generalize to concatenated links with simple quantum memories and active mode switching as in a quantum repeater architecture.  The derivations are technical and are given in the SM, in this section we simply present the resulting expressions for the link efficiency and Bell state fidelity used to model the performance of a concatenated entanglement swapping link with and without the ABSM protocol.  

The efficiency of a $(2N-2)$-fold coincidence yielding a chain of $N-1$ successful BSMs in a passively concatenated link with $N$ independent sources can be written
\begin{align}
\begin{split}
\bar\eta_{1,2N} =\sum_{\vnu\in\{0,1,2\}^N} p^{(\vnu)} \cbeta^{(\vnu)}
\end{split} \label{eq:concateff}
\end{align}
where the sum is taken over all ternary sequences $\vnu=(\nu_1,\nu_2,...,\nu_N)$ in $\{0,1,2\}^N$ representing the number of pairs emitted by each source and 
\begin{equation}\label{eq:pprod}
p^{(\vnu)} = \prod_{i=1}^N p_{2i-1,2i}^{(\nu_i)}.
\end{equation}
The expression \eqref{eq:concateff} arises from a decomposition of the density matrix via Fock space projections in the input modes indexed by $\vnu$.  The coefficients $\cbeta^{(\vnu)}$ can be interpreted as the probability of a full chain of successful BSM outcomes if the $i$-th source produces $\nu_i$ photon pairs.  Note that since we model sources coupled directly to completely dephasing channels, there are no coherent interference effects between states with distinct total photon number from each source (i.e. all off-diagonal density matrix elements between such basis states vanish), and so we may speak unambiguously about the number of photons emitted by each source, subject only to our lack of knowledge of the number of photons emitted in a classical sense (see SM).

The coefficients $\cbeta^{(\vnu)}$ can be written as a product of coefficients $\beta_{ij}^{(m,n)}$ describing the individual success probability of each BSM $B_{ij}$ when $m$ photons are emitted into channel $i$ and $n$ photons are emitted into channel $j$, respectively; however, there is some subtlety to the calculation since neighboring BSMs are correlated by the shared photon source.  These correlations can be accounted by defining the Clifford numbers
\small\allowdisplaybreaks 
\begin{align}\label{eq:errcoeffs1}
\betab_{ij}^{(1,1)} &= \frac{1}{2}\eta_i\eta_j, \\
\betab_{ij}^{(2,0)} &= \frac{\bsigma_{i-1,i}}{3}\eta_i^2, \\
\betab_{ij}^{(0,2)} &= \frac{\bsigma_{j,j+1}}{3}\eta_j^2, \label{eq:beta2002} \\
\betab_{ij}^{(2,1)} &= \eta_i\eta_j(1-\eta_i) + \frac{\bsigma_{i-1,i}}{3}\eta_i^2(1-\eta_j), \label{eq:beta21}\\
\betab_{ij}^{(1,2)} &= \eta_i\eta_j(1-\eta_j) + \frac{\bsigma_{j,j+1}}{3}\eta_j^2(1-\eta_i), \label{eq:beta12}\\
\begin{split}
\betab_{ij}^{(2,2)} &= 2\eta_i\eta_j(1-\eta_i)(1-\eta_j) \\
&\quad + \frac{\bsigma_{i-1,i}}{3}\eta_i^2(1-\eta_j)^2 + \frac{\bsigma_{j,j+1}}{3}\eta_j^2(1-\eta_i)^2, \label{eq:beta22} \\
\betab_{ij}^{(0,0)}&=\betab_{ij}^{(0,1)}=\betab_{ij}^{(1,0)}=0,
\end{split}
\end{align}
\normalsize
where we have introduced abstract vector quantities $\bsigma_{kl}$ associated to each source $S_{kl}$ and perform calculations in the real commutative algebra $\mathcal A$ defined by the relations $\bsigma_{ij}\bsigma_{kl}=\bsigma_{kl}\bsigma_{ij}$ and $\bsigma_{kl}^2 = 3$ for the standard BSM protocol, or $\bsigma_{kl}^2 = 0$ for the ABSM protocol.  Note that real coefficients $\beta_{ij}^{(m,n)}=\mathcal L\betab_{ij}^{(m,n)}$ describing the individual success probability of the BSM $B_{ij}$ are obtained via the linear map $\mathcal L:\mathcal A \to \mathbb R$ defined by $\bsigma_{kl}\mapsto 1$.  Formally, $\mathcal A$ may be realized as an even subalgebra of a Clifford algebra with basis $\{\bsigma_1,...,\bsigma_{2N}\}$.  Namely, $\mathcal A$ is the subalgebra generated by the bivectors $\bsigma_{ij}=\bsigma_i\bsigma_j$ associated to each source $S_{ij}$.

The coefficients defining the efficiency of the concatenated link are then given by
\begin{equation}\label{eq:errcoeff}
\cbeta^{(\vnu)} = \mathcal L\prod_{i=1}^{N-1} \betab_{2i,2i+1}^{(\nu_{i},\nu_{i+1})}
\end{equation}
where the linear map $\mathcal L$ is applied to the product taken in $\mathcal A$.
The relations for $\bsigma_{ij}^2$ can be understood as enforcing the correlated probability for simultaneous measurement of oppositely polarized photons from a single source in both channels $i$ and $j$ adjacent to source $S_{ij}$, where the two distinct relations correspond to the choice of polarization bases used in channels $i$ and $j$.  It should be emphasized that this conditional only holds if both photons of a double-pair are detected---leading to the absence of the $\bsigma_{ij}$ coefficients in \eqref{eq:beta21}, \eqref{eq:beta12}, and \eqref{eq:beta22} on the terms associated to the loss of at least one of the photons in the double-pair.

\subsection{Efficiency and fidelity of a terminated link}
In order to obtain a Bell state with high fidelity when employing probabilistic sources, it is necessary to employ some type of vacuum filtering at the receivers $A$ and $B$ to remove the vacuum component of the state produced by the sources adjacent to the receivers.  This could be in the form of a heralded memory or immediate detection of the distributed Bell state.  Thus, in the following analyses we assume that the concatenated link is terminated by receivers with vacuum filtering.  In the absence of extrinsic noise---employing our simplifying assumption that the outer channels are lossless---the efficiency of a passively concatenated terminated link is given by
\begin{equation}
\bar\eta_{AB}=\sum_{\substack{\vnu\in\{0,1,2\}^N \\ \nu_1,\nu_N>0}} p^{(\vnu)} \cbeta^{(\vnu)}.
\label{eq:passivelinkeff}
\end{equation}
where we omit sequences $\vec \nu$ with $\nu_1=0$ or $\nu_{N}=0$.

The Bell state fidelity for the terminated link is
\begin{equation}\label{eq:fidelity}
F = \frac{\eta_{AB}}{\bar\eta_{AB}}
\end{equation}
where $\bar\eta_{AB}$ appears as the normalization of the state after a successfully terminated BSM chain, and $\eta_{AB}$ is the trace of the projection onto the desired Bell state after the projective measurement of the BSM chain.  The latter can be written (see SM)
\begin{equation}
\eta_{AB}=\sum_{\substack{\vnu\in\{0,1,2\}^N \\ \nu_1=\nu_N=1}} p^{(\vnu)}\Big(\frac{1}{4} \cbeta^{(\vnu)} + \frac{3}{4}\Big[\frac{2}{3}\Big]^{n_2} \hat\beta^{(\vnu)}\Big) \label{eq:concatefftrue}
\end{equation}
where $n_2$ is the number of double-pairs $\nu_i=2$ in $\vnu$ and $\hat\beta^{(\vnu)}$ is the probability of a chain of successful BSMs triggered by exactly one photon from each adjacent source, assuming the corresponding photon numbers 2$\vnu$ emitted by each source.  The coefficient $\hat\beta^{(\vnu)}$ is given by a product of the form $\eqref{eq:errcoeff}$ except that the Clifford product is replaced by a product of real numbers $\hat\beta_{ij}^{(\nu_i,\nu_j)}$ obtained from $\betab_{ij}^{(\nu_i,\nu_j)}$ by substituting $\bsigma_{kl}=0$ to eliminate all terms where two photons are detected from one source.

A derivation of $\eqref{eq:concatefftrue}$ is given in the SM; however, we note here that the first term can be understood as the minimum contribution to the Bell state fidelity of 1/4 corresponding to a completely mixed 2-photon state shared by $A/B$.  The second term represents an additional contribution if the chain of BSMs is unbroken (i.e. if one photon is detected from each adjacent channel).  This additional contribution is reduced by a factor 2/3 for each double-pair, corresponding to the 1/3 probability of the mixed four-photon term in \eqref{eq:4photonstate}.  As a corollary of the result \eqref{eq:concatefftrue}, we find that the fidelity of the Bell state produced by a concatenated entanglement swap in the case that all of the secondary photon emissions are lost is
\begin{equation}\label{eq:dpfidelity}
F^{(\vnu)} = \frac{1}{4} + \frac{3}{4}\Big(\frac{2}{3}\Big)^{n_2}.
\end{equation}

\subsection{Efficiency and fidelity for repeater chains}

For repeater chains supporting multiplexed elementary links, active mode switching allows certain swaps to be performed contingent on the outcome of other BSMs in the chain.
In this case, the post-measurement normalization of the state $\bar\eta_{AB}$ does not directly determine the efficiency $\tilde\eta_{AB}$ of the link---defined as the number of entangled pairs produced per source mode---which depends on the BSM ordering scheme \cite{SHCHUKIN2022,SANGOUARD2011}.  In this work, we restrict our attention to a two-level repeater scheme where all repeater node BSMs are performed simultaneously, independent of the results at neighboring repeater nodes.  This assumption is made primarily to simplify the analysis, but it also has the benefit of placing a minimal requirement on the lifetime of the quantum memories employed in the repeaters.  

Treating each elementary link connecting sources $S_{i-3,i-2}$ and $S_{i-1,i}$ as a Bernoulli trial with success probability $\bar\eta_{i-3,i}$, the average number of source modes (mode-pairs for dual-rail qubits) $\mu_1$ needed per elementary link in order to establish a successful BSM in all $\ell$ elementary links is given by 
\begin{equation}
\mu_{1} = \frac{1}{\ell}\sum_{i=1}^\ell \frac{1}{\bar\eta_{4i-3,4i}}.
\end{equation}
Assuming a two-level scheme, all of the first-level elementary links are discarded if at least one of the repeater node BSMs fails.  To determine the success probability of the second level of the link (consisting of all of the repeater node BSMs), we can consider an ensemble of attempts to fully connect a passively concatenated link.  Since the sources are all independent, the fraction of attempts in which all of the elementary links succeed is given by $P_1=\bar\eta_{14}\bar\eta_{58}\cdots \bar\eta_{4\ell-3,4\ell}$.  The probability that the second level of the link succeeds is the relative fraction of this subset of attempts for which the full passive link succeeds.  The full terminated passive link is connected with probability $\bar\eta_{AB}$, hence, the probability that the second level of the repeater link succeeds is
$$P_2 = \frac{\bar\eta_{AB}}{P_1}= \frac{\bar\eta_{AB}}{\bar\eta_{14}\bar\eta_{58}\cdots \bar\eta_{4\ell-3,4\ell}}.$$
Since it takes on average $1/P_2$ attempts for the second level of the link to succeed, the average number of source modes needed to obtain a fully connected chain is given by $\mu_2 = (1/P_2)\mu_1$.  Thus, the number of entangled pairs produced per source mode is given by
\begin{equation}
\tilde \eta_{AB} = \frac{1}{\mu_2} = \ell \frac{\bar\eta_{AB}}{\prod_{i=1}^\ell \bar\eta_{4i-3,4i} \sum_{j=1}^\ell \bar\eta_{4j-3,4j}^{-1}}.
\end{equation}
Assuming that every source node $S_{ij}$ produces pairs at a multiplexed rate weighted by the inverse efficiency $\bar\eta_{k-3,k}^{-1}$ of the adjacent elementary link (to avoid bottle-necking the chain), the overall rate at which entangled pairs are produced by the chain is given by
\begin{equation}
R_{AB} = R \tilde\eta_{AB}
\end{equation}
where $R$ is the average multiplexed pair rate employed by a single source node in the chain.  The efficiency $\tilde \eta_{AB}$ is the relevant figure of merit for the repeater chain in that it determines the overall entangled pair rate up to the average multiplexed rate $R$ of the source nodes.
 
Neglecting infidelities introduced by the quantum memories and mode-switching mechanism (i.e. modeling the memories as equivalent to low-loss optical loops which don't affect the state aside from an overall attenuation included in the repeater channel efficiency), the state delivered to $A/B$ by the repeater chain contingent on success of all internal BSMs is identical to the state supplied by passively concatenated links and is thus also given by \eqref{eq:concateff}, \eqref{eq:fidelity}, and \eqref{eq:concatefftrue} (or $\eqref{eq:passivelinkeff}$-$\eqref{eq:concatefftrue}$ for a terminated link).  This follows from the fact that each source is independent and all of the generalized measurement operators associated to different BSMs commute.  Thus, the effect of active switching between multiplexed mode-pairs available at each repeater node is to more efficiently produce out of an ensemble of equivalent source states a pair of modes at $A$ and $B$ which have been connected (post-measurement) by a chain of successful BSMs.

\section{Multi-photon noise in quantum repeater chains}\label{sec:repeaters}

\begin{figure}[h!]
\centering\includegraphics[width=8.7cm]{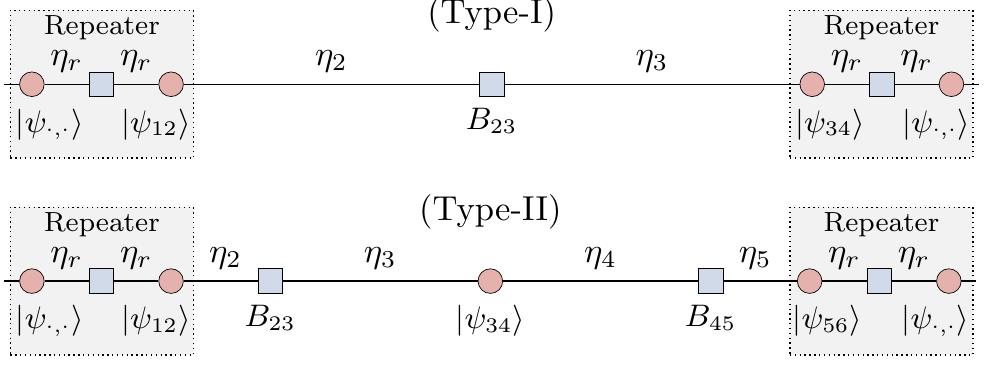}
\captionsetup{font=footnotesize,labelfont=footnotesize,justification=raggedright}
\caption{Two types of elementary links connecting repeater nodes.  The Type-I link is a typical architecture for a terrestrial quantum network.  The Type-II link is motivated by satellite-assisted entanglement distribution, where an entangled pair source $S_{34}$ distributes entangled pairs $|\psi_{34}\rangle$ to adjacent repeater stations from an orbiting satellite.}
\label{fig:linktypes}
\end{figure}

In this section we quantify the suppression of multi-photon noise achieved by the ABSM protocol in quantum repeater chains.  We consider repeater chains connected by two different types of elementary links shown in Figure \ref{fig:linktypes}---referred to below as Type-I and Type-II links.  The Type-I link is a standard entanglement distribution architecture for terrestrial quantum networks (or a satellite uplink) and consists of a single BSM node connecting adjacent repeater nodes.  The Type-II link is motivated by entanglement distribution via a satellite downlink, where an entangled pair source $S_{34}$ distributes entangled photons to adjacent repeater stations from an orbiting satellite.  In the absence of heralded quantum memories which can efficiently capture relatively broadband photons from a lossy downlink, BSMs $B_{23}$ and $B_{45}$ may be employed to verify transmission of the photon without destroying the entangled state.  As illustrated in Figure \ref{fig:linktypes}, we shall consider elementary links with general channel efficiencies $\eta_i$ connecting each source node to the neighboring BSM, with the efficiency of the channels connecting the sources to the repeater node BSMs denoted $\eta_r$.

\subsection{Balanced repeater chains}
\begin{figure}[h!]
\centering\includegraphics[width=8.7cm]{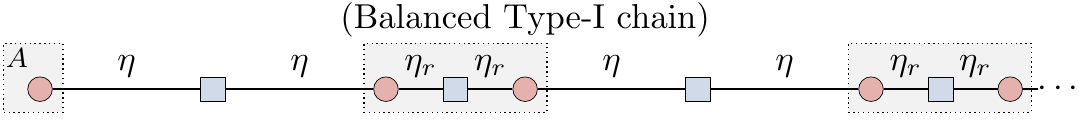}
\captionsetup{font=footnotesize,labelfont=footnotesize,justification=raggedright}
\caption{Repeater chain of balanced Type-I elementary links.}
\label{fig:balancedchain}
\end{figure}

First, we consider a quantum repeater chain consisting of balanced Type-I links with identical channel efficiency $\eta_i\equiv \eta$ adjacent to each BSM and repeater internal efficiency $\eta_r$ (Fig. \ref{fig:balancedchain}).  
The Bell state fidelity for balanced repeater chains with an arbitrary number of elementary links $\ell$ is approximated by considering only multi-pair emission sequences in \eqref{eq:passivelinkeff} with $\vec\nu$ containing at most $2\ell+1$ photons from $N=2\ell$ sources.  These can be enumerated based on the number of vacuum emissions $n_0$ in the sequence
\begin{align*}
(2,1,1,...,1), (1,2,1,1,...,1),...,  &\quad (n_0 = 0), \\
(2,0,2,1,1,...,1), (1,2,0,2,1,1,...,1), ..., &\quad (n_0 = 1), \\
(2,0,2,0,2,1,...,1), (1,2,0,2,0,2,1,...,1), ..., &\quad (n_0 = 2),
\end{align*}
and so forth for $n_0>2$.  Note that there are $2(\ell-n_0)$ terms for each $n_0<\ell$.  The ABSM protocol is such that a single source cannot trigger adjacent BSMs, hence, it completely eliminates all errors with $n_0>1$.  Since the number of error terms with $n_0=0$ and $n_0=1$ grow linearly with the length of the chain $\ell$, while the remaining errors with $n_0>1$ grow quadratically in $\ell$, the ABSM protocol completely suppresses the dominant quadratic growth of multi-photon errors for long repeater chains.

To quantify this we use the expressions of Section \ref{sec:linkmodel} to derive the Bell state fidelity for a balanced repeater chain of arbitrary length.  Remarkably, using the Clifford product the sums in \eqref{eq:passivelinkeff}-\eqref{eq:concatefftrue} can be evaluated in closed form yielding a compact expression for the fidelity
\begin{equation}\label{eq:bsfidelitychain}
F = \frac{1 + p\big(\varepsilon_+ + \frac{11+\sigma}{5+\sigma} \varepsilon_0\big) + O(p^2)}{1 + 4 p (\varepsilon_0 + \varepsilon_0' + \varepsilon_+ + \varepsilon_+') + O(p^2)},
\end{equation}
where $p\equiv p_{ij}$ is the source emission probability, the terms $\varepsilon_0'$,$\varepsilon_0$ give the contribution from $n_0=0$ errors
\begin{align}
\varepsilon_0' &= (1-\eta) \\
\varepsilon_0 &= \frac{\ell - 1}{2} (5 + \sigma)(1-\eta)(1-\eta_r),
\end{align}
and $\varepsilon_+'$, $\varepsilon_+$ represent the remaining sum for $n_0>0$, taking the value $\varepsilon_+=\varepsilon_+'=0$ for $\ell=1$ and 
\begin{align}
\varepsilon_+' &= \frac{1}{4}(1+\sigma)[1 + \sigma(\ell-2)](1-\eta) \\
\varepsilon_+ &= \frac{\ell-2}{4} (1 + \sigma)^2 [2 + \sigma(\ell - 3)](1-\eta)(1-\eta_r) \label{eq:gammap}
\end{align}
for $\ell > 1$, where $\sigma=\bsigma_{ij}^2/3$ depends on the BSM protocol.  The coefficient $\sigma$ completely captures the noise suppression provided by the ABSM protocol with $\sigma = 1$ for the non-alternating BSM protocol and $\sigma = 0$ for the ABSM protocol.  The terms $\varepsilon_0'$ and $\varepsilon_+'$ represent the contribution from boundary sequences with multi-pair emissions $\nu_1=2$ or $\nu_N=2$.

\begin{figure}[h!]
\centering\includegraphics[width=8.7cm]{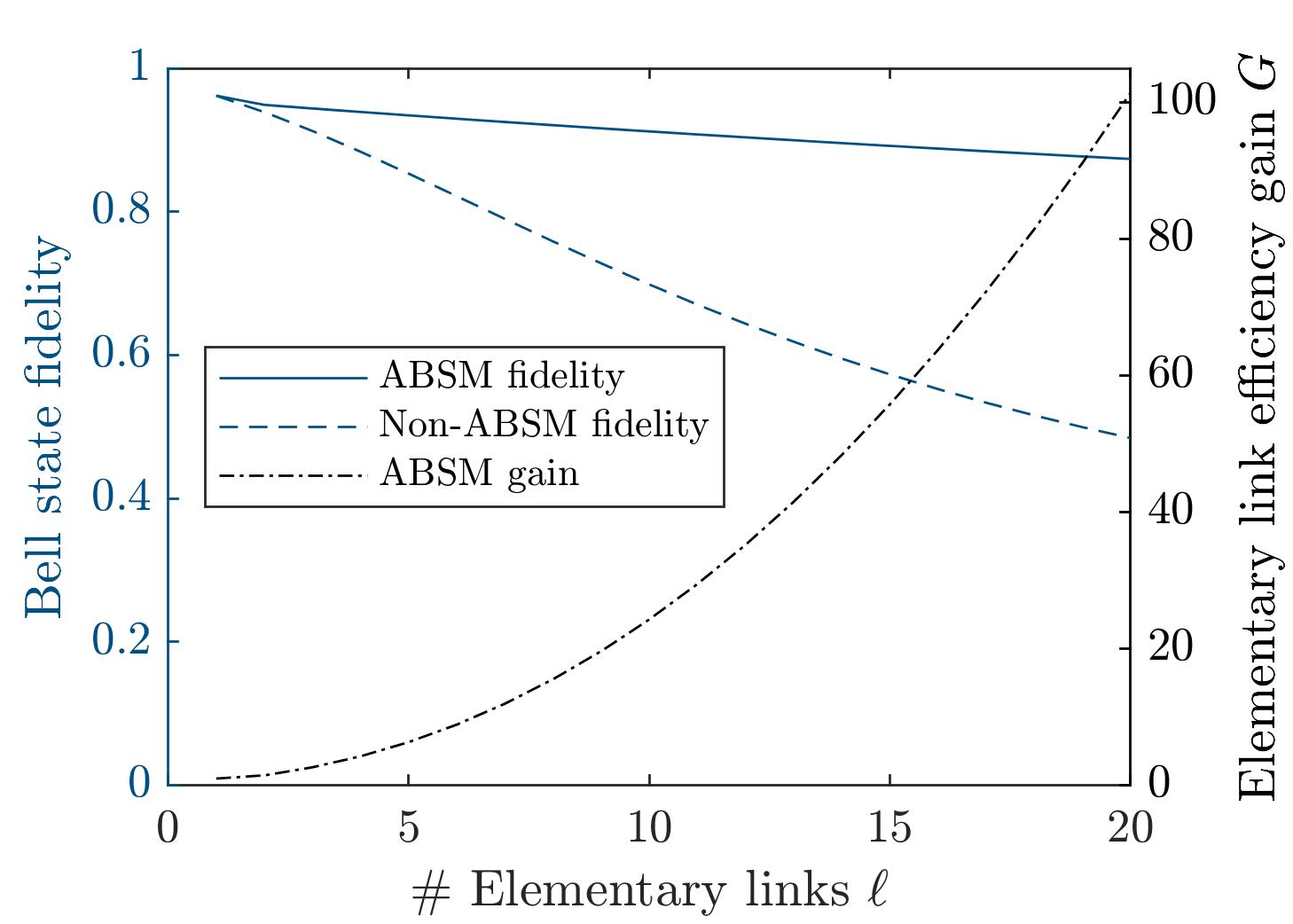}
\captionsetup{font=footnotesize,labelfont=footnotesize,justification=raggedright}
\caption{Bell state fidelity for repeater chains using probabilistic sources with emission probability $p=0.01$ and repeater internal efficiency $\eta_r = 0.9$.  The axis at right shows the gain in elementary link efficiency (proportional to $p^2$) provided by the ABSM protocol if the emission probabilities are adjusted to achieve a Bell state fidelity $F=0.9$ with $\eta_r=0.9$.}
\label{fig:repeatergain}
\end{figure}

The last term \eqref{eq:gammap} quantifies the dominant $\ell^2$ growth of multi-photon errors in \eqref{eq:bsfidelitychain} when $\sigma=1$; however, this quadratic term vanishes when employing the ABSM protocol ($\sigma=0$).  Note that the fidelity is essentially independent of the channel efficiency $\eta$ for $\eta\ll 1$ since we can use the approximation $(1-\eta)\simeq 1$.  The Bell state fidelity for the balanced chain thus depends only on the repeater efficiency $\eta_r$ and the number of elementary links $\ell$.  The dependence of the Bell state fidelity on the length of the chain is shown in Figure \ref{fig:repeatergain} for repeaters with $\eta_r = 0.9$.

Note the significant improvement in fidelity provided by the ABSM protocol as the length of the chain increases.  The impact of this on repeater chains employing probabilistic sources can be translated into a gain $G$ in efficiency for each elementary link subject to a fixed fidelity $F$ by inverting \eqref{eq:bsfidelitychain} for the allowed emission probability
\begin{align}\label{eq:pp}
\begin{split}
p(F;\sigma) =\frac{1-F}{4F[\varepsilon_0 + \varepsilon_0' + \varepsilon_+ + \varepsilon_+'] - \varepsilon_+ - \frac{11+\sigma}{5+\sigma} \varepsilon_0}.
\end{split}
\end{align}
The elementary link efficiency depends on the square of the emission probability $p$, and so the gain is given by
\begin{equation}\label{eq:repeatergain}
G = \frac{p(F;\sigma = 0)^2}{p(F;\sigma = 1)^2}.
\end{equation}
The result is shown in Figure \ref{fig:repeatergain} for chains with $\eta_r=0.9$, assuming fixed fidelity $F=0.9$.  From \eqref{eq:pp}-\eqref{eq:repeatergain} we see that the gain grows quadratically with the length of the chain, reaching a 20-fold increase for a chain with $\ell=10$ elementary links. This quadratic growth can again be understood in correspondence with our earlier observation of the suppression of errors with $n_0>1$.

\begin{figure}[h!]
\centering\includegraphics[width=8.7cm]{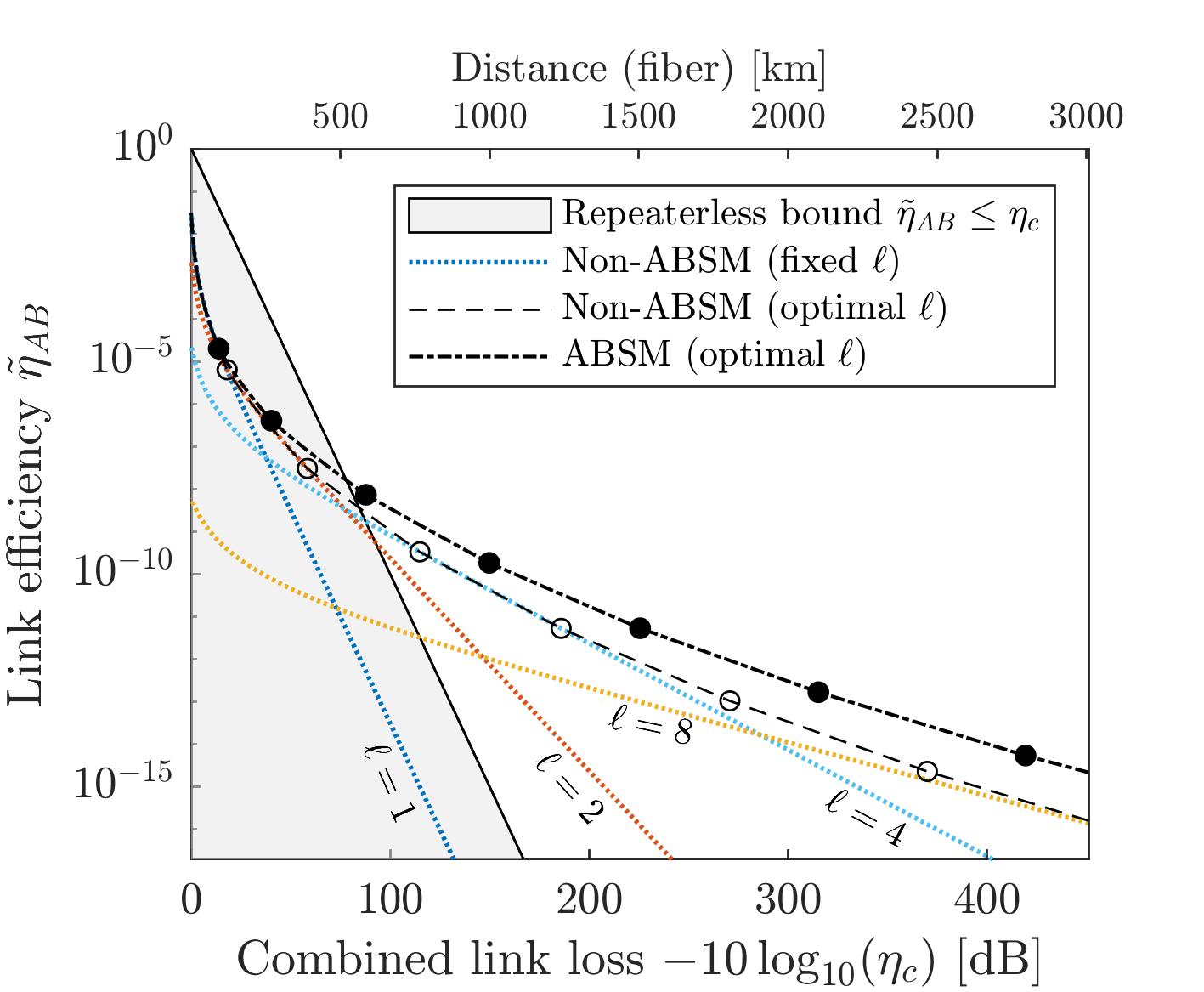}
\captionsetup{font=footnotesize,labelfont=footnotesize,justification=raggedright}
\caption{Link efficiency as a function of total link loss for a repeater chain using PDC sources with pair probability configured to yield a Bell state fidelity $F=0.9$ with $\eta_r=\eta_d=0.9$.  The dotted curves show the efficiency without the ABSM protocol for $\ell = 1$, 2, 4, and 8.  The dashed curves show the maximum efficiency optimizing the number of links $\ell$ with and without the ABSM protocol.  The markers show where the optimal number of links increases.  The shaded region shows the repeaterless bound $\tilde \eta_{AB}\leq \eta_c$ for perfect deterministic sources.}
\label{fig:repeaterefficiency}
\end{figure}

Employing the two-level repeater model presented in Section \ref{sec:linkmodel}, the overall link efficiency for a chain with total combined channel loss $\eta_{c}$ divided into $\ell$ elementary links is
\begin{equation}
\tilde \eta_{AB} = \frac{\bar \eta_{AB}}{\bar\eta_{\textup{EL}}^{\ell - 1}} = p^{2}\eta^{2} \frac{\eta_r^{2(\ell-1)}}{2^{2\ell-1}} + O(p^3)
\end{equation}
where $\eta = \eta_c^{1/2\ell}\eta_d$, $\eta_d$ is the efficiency of the detectors used for the linear optical BSMs at the center of the elementary links, and
\begin{equation}
\bar\eta_{\textup{EL}} \equiv \bar\eta_{14}=\bar\eta_{58}=...=\bar\eta_{4\ell-3,4\ell} = p^2\eta^2 + O(p^3)
\end{equation}
is the elementary link efficiency for the balanced chain (including double-pair errors).  To lowest order in the pair probability $p$, the result is the elementary link efficiency $\bar\eta_{\textup{EL}}$ reduced by the factor $\eta_r^{2(\ell-1)}/2^{2\ell-1}$ arising from the repeater efficiency and 50\% success probability of the $2\ell-1$ linear optical BSMs (here $\eta_r$ should be understood to include the detection efficiency of the repeater node BSM).  Figure \ref{fig:repeaterefficiency} shows the link efficiency as a function of the total combined loss $\eta_c$, translated into a terrestrial link distance by assuming a fiber attenuation of $0.15$ dB/km.

Note that even without the ABSM protocol, the repeater chain with PNR BSMs can outperform the repeaterless bound $\tilde\eta_{AB} \leq \eta_c$ for distribution of entangled photon pairs using a perfect deterministic pair source.  More precisely, assuming repeaters with internal efficiency $\eta_r= 0.9$ the repeater architecture analyzed here first surpasses this ideal repeaterless link at fidelity $F=0.9$ with an optimal number $\ell = 3$ elementary links dividing a combined loss of 83 dB.  However, in order to achieve a rate of $10^5$ pairs/s at this range requires a multiplexed repetition rate $\sim$10$^{13}$ Hz from each source.  This result is very similar the multiplexed rate of $3\cdot 10^{13}$ Hz found to surpass an ideal repeaterless QKD bound at a rate of $\sim$10$^5$ secret key bits/s in a similar PDC repeater architecture studied in \cite{KROVI2016}.  In that work, an ideal repeaterless bound was also found to be first surpassed using 3 elementary links to divide a combined loss of $\sim$90 dB assuming sources with $10^4$ frequency modes and 100 spatial modes operated at a repetition rate of 30 MHz.

The upper envelope for the link efficiency---optimizing the number of links $\ell$ for a given total link loss---is shown in Figure \ref{fig:repeaterefficiency} for both the standard protocol and the ABSM protocol.  As the link efficiency $\tilde\eta_{AB}$ is proportional to the elementary link efficiency $\bar\eta_{\textup{EL}}$ and determines the entangled pair rate up to the multiplexed rate $R$ of each source, the gain in the entangled pair distribution rate provided by the ABSM protocol is in direct correspondence with the elementary link gain shown in Figure \ref{fig:repeatergain} and reaches a factor of 10 for chains with $\ell = 6$ elementary links.

\subsection{Noise suppression for elementary links with imbalanced losses}
The results of the previous section show that using PDC-based repeaters, very high multiplexed source rates are required to achieve appreciable entangled pair rates at 100+ dB combined link loss.  Thus, even with repeaters, long distance links based on PDC sources demand a more efficient conversion from combined link loss to link distance than is provided by the $\sim$0.15 dB/km from fiber transmission loss.  Satellite-assisted links governed by free-space diffraction loss can make more effective use of the allowable total link loss to achieve longer range links; however, the architecture of satellite-assisted links restricts the location of BSM nodes leading to dynamic, imbalanced channels which can be detrimental to PDC-based links with multi-photon noise.  

\begin{figure}[h!]
\centering\includegraphics[width=8.7cm]{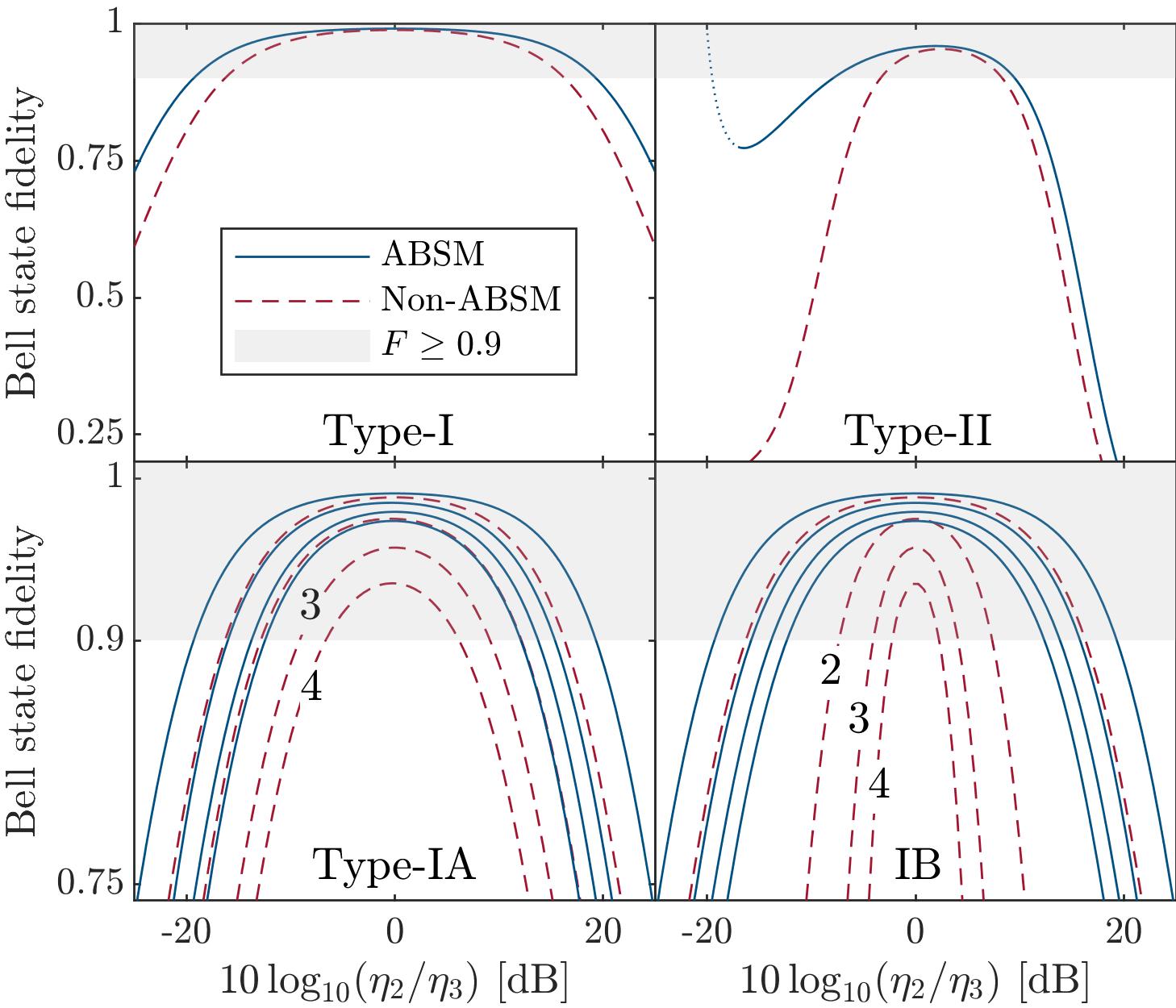}
\captionsetup{font=footnotesize,labelfont=footnotesize,justification=raggedright}
\caption{The top two panels show the Bell state fidelity for a single elementary link (\emph{cf}. Fig. \ref{fig:linktypes}) with imbalanced channels dividing 40 dB combined loss with pair probability $p=0.01$.  The bottom two panels show the fidelity for Type-I repeater chains with $\ell=1,2,3,$ and $4$ elementary links (from top to bottom).  The Type-II link assumes a symmetric imbalance $\eta_2=\eta_5$ and $\eta_3=\eta_4$; the dotted curve denotes the domain where 3-pair emissions---neglected in this analysis---become the dominant source of noise.}
\label{fig:bellstatefidelity}
\end{figure}

Motivated by this problem, we analyze the additional suppression of multi-photon noise provided by the ABSM protocol in chains with imbalanced channel losses.  To quantify this, we first consider the reduction in fidelity caused by imbalanced losses which are not compensated by a corresponding suppression of single-pair emissions---an analysis of the gain provided by the protocol if the imbalanced losses are compensated is given in Section \ref{sec:absmgain}.  Figure \ref{fig:bellstatefidelity} shows the dependence of the fidelity on channel losses for elementary links with 40 dB combined loss, assuming the source emission probabilities are all fixed at $p_{ij}=0.01$ and the multi-pair emissions are governed by \eqref{eq:npaird}.  An efficiency $\eta_r=0.9$ is assumed for all channels internal to the quantum repeaters.

The top left panel shows that the Type-I link fidelity is symmetric in the link imbalance, and the ABSM protocol supports an additional 3 dB difference in channel losses at 90\% state fidelity.  The top right panel shows the Type-II link fidelity for symmetric imbalances $\eta_2=\eta_5$ and $\eta_3=\eta_4$ (\emph{cf.} Fig. \ref{fig:linktypes}).  The ABSM protocol significantly suppresses multi-photon noise for Type-II links such that both BSMs are located closer to the central source.  In this case, the ABSM fidelity remains above 74\%, supporting an additional 5 dB imbalance at 90\% state fidelity.  On the other hand, if both BSMs are located closer to the repeater nodes---as in the satellite downlink architecture described above---the ABSM protocol only supports an additional 1-2 dB difference in channel losses at 90\% state fidelity.  The fidelity generally declines with larger differences in channel loss.

As the length of the repeater chain grows, the reduction in fidelity is compounded as shown in the previous section for balanced links \eqref{eq:bsfidelitychain}-\eqref{eq:gammap}.  This reduction is even more severe in the presence of imbalanced channels; however, the bottom two panels in Figure \ref{fig:bellstatefidelity} show that the ABSM protocol is more resilient to this reduction in fidelity for longer chains of Type-I links.  

\begin{figure}[h!]
\centering\includegraphics[width=8.7cm]{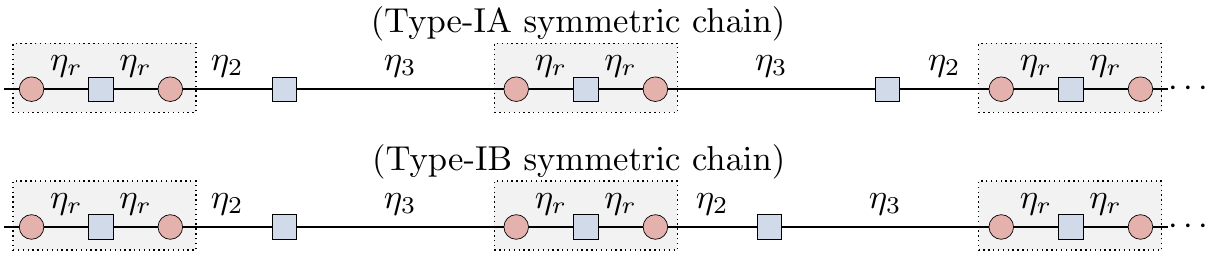}
\captionsetup{font=footnotesize,labelfont=footnotesize,justification=raggedright}
\caption{Repeater chains of imbalanced Type-I elementary links with special symmetries which demonstrate the non-locality of multi-photon noise in extended repeater chains.}
\label{fig:imbalancedchains}
\end{figure}

Figure \ref{fig:bellstatefidelity} also shows that the severity of multi-photon noise from chains of imbalanced links can depend significantly on the direction in which each elementary link is imbalanced.  The dependence of the fidelity on the link imbalance is shown for chains of imbalanced Type-I links with two special symmetry types, labeled Type-IA and Type-IB (Figure \ref{fig:imbalancedchains}).
The Type-IA chain is defined with the links imbalanced in alternating directions, so that $\eta_2\leftrightarrow\eta_3$ are exchanged for adjacent elementary links.  The Type-IB chain is defined such that all of the links are imbalanced in the same direction (and by the same amount).  Note that the Bell state fidelity degrades much more quickly in the Type-IB chain using the standard BSM protocol.  This behavior shows that multi-photon noise in a quantum repeater chain cannot be considered purely locally, i.e., noise can propagate down the chain.

To understand this phenomenon, we must consider the dominant source of multi-photon noise in each chain.  In the Type-IB chain, the dominant noise in \eqref{eq:passivelinkeff} is given by double pairs separated by a single intermediate source.  These can be connected through successful BSMs by subsequences $\vnu$ of the form $(2,0,2,0,...,2)$ consisting of only $N+1$ photon pair emissions from $N$ sources.  Conversely, the dominant source of multi-photon noise in the Type-IA chain comes from double pairs separated by an even number of intermediate sources, e.g. $(2,2)$ or $(2,1,1,2)$, which require at least $N+2$ photon pair emissions from $N$ sources and are thus not as prominent.  In this way, we see that the repeater nodes do not insulate multi-photon errors from neighboring links, i.e., the final Bell state fidelity cannot generally be determined by combining an independent reduction in fidelity from each elementary link.

Indeed, this lack of independence of the multi-pair noise is precisely the property that the ABSM protocol takes advantage of, since multi-pair events are suppressed based on their correlation to multi-pair events in neighboring links.  In doing so, the ABSM protocol can be seen to block the propagation of these errors.  Specifically, the ABSM protocol yields essentially the same fidelity for both Type-IA and Type-IB chains.  This can be understood by the observation that the ABSM protocol does not allow a BSM chain to be connected by emission sequences of the form $(0,2,0)$, which produce the dominant error in the Type-IB chain.

\section{Multi-photon noise in Type-II elementary links}\label{sec:absmgain}

In this section we consider an application of the ABSM protocol to elementary links consisting of two simultaneous entanglement swaps (Type-II links).  This type of architecture is motivated by a satellite downlink and avoids many engineering challenges associated to establishing a synchronized, low-loss optical uplink through Earth's atmosphere.  The idea of the Type-II link is to use an entanglement swap with a high-rate probabilistic photon source as a quantum non-demolition (QND) measurement of the presence of another photonic entangled state before or after a lossy entanglement distribution link.  For example, in the absence of heralded quantum memories which can efficiently capture relatively broadband photons from a lossy downlink, one can instead position a PDC source in a ground station receiving a satellite downlink to verify transmission of a photon using a local BSM.

\begin{figure}[h!]
\centering\includegraphics[width=8.7cm]{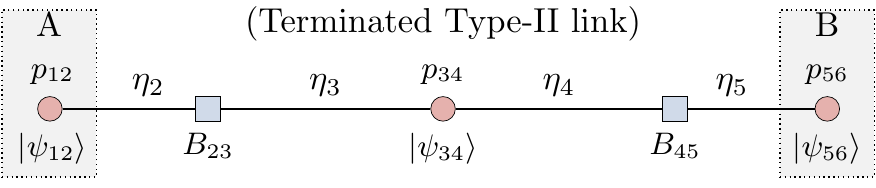}
\captionsetup{font=footnotesize,labelfont=footnotesize,justification=raggedright}
\caption{A terminated Type-II link motivated by satellite-assisted entanglement distribution.  The efficiency of the link with PDC sources and imbalanced losses $\eta_i$ depends on the maximum allowed pair probabilities $p_{12},p_{34},p_{56}$ subject to a constraint on the minimum Bell state fidelity.}
\label{fig:type2link}
\end{figure}

One of the drawbacks of using a PDC source in this manner is that if a BSM is not centrally located between two PDC sources, one source can dominate the BSM with double-pairs relative to the two-photon coincidences with one photon from each side.  This is detrimental to the efficiency of the link, since it requires that the link be artificially balanced by reducing the emission probability of the source close to the BSM to suppress double-pairs.  In a double entanglement swap the ABSM protocol can identify some of the double-pair errors potentially allowing for a higher emission probability.  Double-pairs can be further suppressed by combining the ABSM protocol with a `cascaded' entangled pair source as described in \cite{DHARA2022}---which employs a high-efficiency BSM with PNR to herald the production of a single entangled pair.  Before proceeding to an analysis of entanglement distribution with cascaded PDC sources, we first demonstrate the principle by quantifying the gain afforded by the ABSM protocol for a terminated elementary link consisting of two passively concatenated entanglement swapping links with 3 independent sources (Figure \ref{fig:type2link}).

\subsection{Maximum efficiency for double entanglement swap with probabilistic sources and imbalanced links} 
In the following analysis, we assume that the double entanglement swapping link is passively concatenated (i.e. both BSMs must succeed simultaneously for a successful entanglement distribution).  Following the model of a terminated link presented in Section \ref{sec:linkmodel}, the probability of a four-fold coincidence with opposite polarizations in each BSM and at least one photon sent to $A$ and $B$ during a single use of the link takes the form
\small
\begin{align}
\begin{split}
&\bar\eta_{AB} = \sum_{\substack{i,j,k=0\\ i,j>0}}^2 p_{12}^{(i)}p_{34}^{(j)}p_{56}^{(k)} \cbeta^{(i,j,k)}
\end{split}
\end{align}
\normalsize
where the coefficient $\cbeta^{(l,m,n)}$ describes the probability of a four-fold coincidence associated to an $l$-pair emission from source $S_{12}$, $m$-pair emission from source $S_{34}$, and $n$-pair emission from $S_{56}$.  We neglect events with more than 4 total photon pairs produced by the 3 sources.

The primary coefficient is the (1,1,1)-coefficient describing the probability of two simultaneously successful BSMs given a single-pair emission from each source
\begin{align}
\cbeta^{(1,1,1)} &= \frac{1}{4}\eta_2\eta_3\eta_4\eta_5.
\end{align}
The leading order error coefficients are then calculated explicitly from Clifford products of \eqref{eq:beta2002}-\eqref{eq:beta12} as
\small
\begin{align}\label{eq:loerr1}
\hat\beta^{(1,2,1)} &= \eta_2\eta_3\eta_4\eta_5(1-\eta_3)(1-\eta_4)  \\
\begin{split}
\cbeta^{(1,2,1)} &= \hat\beta^{(1,2,1)} + \frac{\sigma}{3}\eta_3^2\eta_4^2(1-\eta_2)(1-\eta_5)  \\
+\frac{1}{3} \eta_2 & \eta_3\eta_4^2(1-\eta_3)(1-\eta_5) + \frac{1}{3} \eta_3^2 \eta_4\eta_5(1-\eta_2)(1-\eta_4)  \label{eq:loerr3}
\end{split} \\
\cbeta^{(2,1,1)} &= \frac{1}{2}\eta_2\eta_3\eta_4\eta_5(1-\eta_2) + \frac{1}{6}\eta_2^2\eta_4\eta_5(1-\eta_3) \\
\cbeta^{(2,0,2)} &= \frac{1}{9}\eta_2^2\eta_5^2
\end{align}
\normalsize
with the (1,1,2) coefficient related to the (2,1,1) coefficient via the index substitution $2\leftrightarrow 5$ and $3\leftrightarrow 4$.
The parameter $\sigma = \bsigma_{34}^2/3$ again captures the suppression of double-pair emission noise achieved using an alternating basis in the BSMs, with $\sigma=0$ for the ABSM protocol and $\sigma=1$ using non-alternating BSMs.  Note that this suppression is most significant when link imbalances are such that $\eta_3^2\eta_4^2$ is the dominant term.  The last two coefficients above are unaffected by the protocol since they do not include events where the source adjacent to both BSMs produces a double-pair; however, as discussed at the end of the section these noise events can be suppressed by QND measurements with PNR at the receivers $A$ and $B$ since double-pair states are produced at an outer node.

For the double-swap link, the Bell state fidelity is
\begin{equation}
F = p_{12}p_{56}\frac{p_{34}\beta^{(1,1,1)} + p_{34}^{(2)}[\frac{1}{4}\beta^{(1,2,1)} + \frac{1}{2}\hat\beta^{(1,2,1)}]}{\bar\eta_{AB}}.
\end{equation}
To obtain analytical results, we shall approximate the fidelity $F$ by calculating $\bar\eta_{AB}$ using only the leading order error coefficients given above.

Assuming the multi-pair emissions are governed by \eqref{eq:npaird}, we approximate the relation of single-pair to double-pair emissions as 
\begin{align}
p_{ij}^\2 &\simeq \frac{3}{4}p_{ij}^2.
\end{align}
The impact of multi-pair emission noise for fixed link losses $\eta_i$ is determined by considering the maximum efficiency subject to a fixed infidelity $\Delta f$ 
\begin{equation}
\hat\eta_{AB}(\Delta f) = \max_{p_{12},p_{34},p_{56}}\Big\{\eta_{AB} : 
\begin{array}{ll} F \geq 1-\Delta f\\
 p_{ij}\leq 8/27\end{array} \Big\}
\end{equation}
where the maximum is taken over all emission probabilities satisfying $p_{ij}\leq 8/27\simeq 0.3$ in accord with the maximum allowable single-pair probability in \eqref{eq:npaird}.  

In order to analyze the maximum efficiency we use the method of Lagrange to solve the constrained optimization under the assumption that the optimum $\hat\eta_{AB}(\Delta f)$ is obtained on the boundary $F=1-\Delta f$ with $p_{ij} < 8/27$---conditions that generally hold provided all channels have losses $\eta_i<1$ and $\Delta f$ is sufficiently small.  
The Lagrangian constraint 
leads to a relation defining the relative pair probabilities which maximize the efficiency $\eta_{AB}$ for fixed fidelity (see SM)
\begin{align}\label{eq:relativepp}
\begin{split}
p_{12}b_{12} = p_{34}b_{34}w(b) = p_{56}b_{56} \equiv p.
\end{split}
\end{align}
where the weights are given by
\small
\begin{align}
b_{12} &= \beta^{(2,1,1)},\quad b_{56} = \beta^{(1,1,2)} \\
\quad b_{34}&=\frac{3}{4}\beta^{(1,2,1)}-\frac{1}{2}\hat\beta^{(1,2,1)}
\end{align}
\normalsize
and the central source emission probability has an additional weight $w(b)=2/(1+\sqrt{1+6b})$ associated to balancing the $\cbeta^{(2,0,2)}$ noise determined by the parameter 
\begin{align}\label{eq:bparameter}
\begin{split}
b &= \frac{b_{34}\cbeta^{(2,0,2)}}{b_{12}b_{56}}.
\end{split}
\end{align}
The latter satisfies $(\sigma/36)\leq b \leq 1$ and increases as the BSMs move farther from the central source towards the outer sources.  

Noting that $F$ is a differentiable function of the parameter $p$ with non-vanishing derivative at $p=0$, the linear approximation to $p$ determining the optimal probabilities $p_{ij}$ can be evaluated near $\Delta f=0$
\begin{equation}
p = \frac{16\beta^{(1,1,1)}}{36 + 27bw(b)} \Delta f + O\big((\Delta f)^2\big)
\end{equation}
yielding the maximum efficiency with infidelity $\Delta f$
\small
\begin{equation}\label{eq:maxeff}
\hat\eta_{AB} = \frac{16^3[\cbeta^{(1,1,1)}]^4}{b_{12} b_{34} b_{56} w(b)[36+27bw(b)]^3} (\Delta f)^3 + O\big((\Delta f)^4\big).
\end{equation}
\normalsize

\subsection{Efficiency penalty and ABSM gain for imbalanced links with probabilistic sources}
 
In order to gain insight into the main result \eqref{eq:maxeff} from the previous section, we factor the expression in the form
\begin{equation}\label{eq:factoredeff}
\hat\eta_{AB} = \pi_0 \beta^{(1,1,1)} \hat\pi(\eta_2,\eta_3,\eta_4,\eta_5) + O\big((\Delta f)^4)
\end{equation}
where $\beta^{(1,1,1)}$ represents the efficiency of the link with deterministic entangled pair sources and $\pi_0$ represents the combined source efficiency $p_{12}p_{34}p_{56}$ using probabilistic sources in the limit of balanced low-efficiency links $\eta_2=\eta_3=\eta_4=\eta_5\ll 1$.  The remaining factor $\hat\pi$ captures both the balancing loss associated to a reduction in source efficiency required to compensate imbalanced channels, as well as the gain from PNR in the BSM detectors which discards multi-photon events when at least 3 photons are detected.

Using the results from the previous section, the combined source efficiency in the limit of balanced, lossy channels is given by
\begin{align}\label{eq:pi0doubleswap}
\begin{split}
\pi_0 &= \begin{cases} 0.0027\cdot (\Delta f)^3, & \sigma=1 \\
0.0037 \cdot(\Delta f)^3, & \sigma = 0, \end{cases}
\end{split}
&(\textup{Type-II}).
\end{align}
For comparison, the combined source efficiency $\pi_0$ for a balanced Type-I link can be obtained from \eqref{eq:pp} by setting $\ell=1$, and is given to leading order in $\Delta f$ as
\begin{align}\label{eq:pi0singleswap}
\begin{split}
\pi_0 &= 0.0625 \cdot(\Delta f)^2,
\end{split}
&(\textup{Type-I}).
\end{align} 
Thus, even with balanced losses the Type-II link suffers an additional factor of $0.04\cdot \Delta f$ efficiency reduction relative to the single-swap Type-I link.

\begin{figure}[h!]
\centering\includegraphics[width=8.0cm]{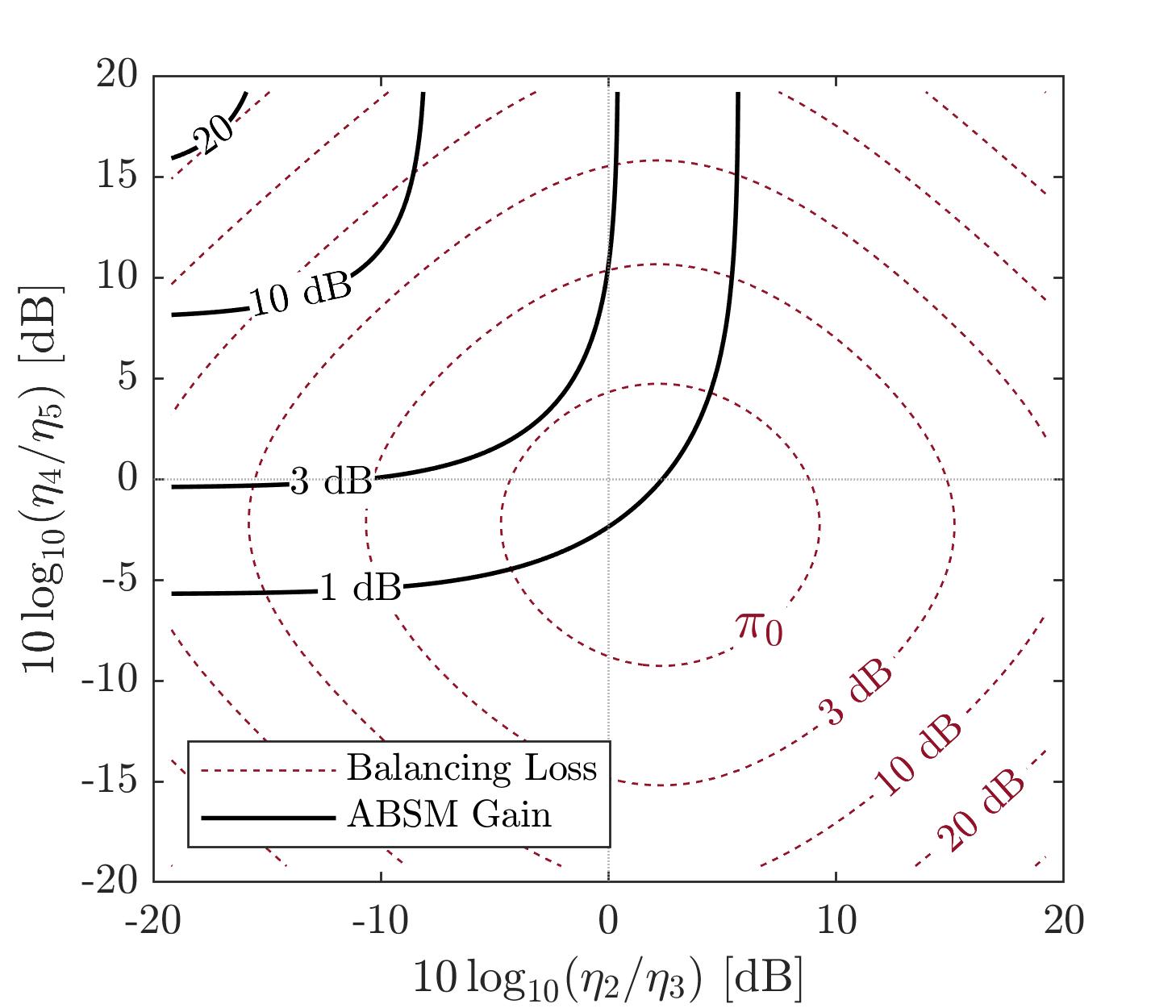}
\captionsetup{font=footnotesize,labelfont=footnotesize,justification=raggedright}
\caption{Gain in link efficiency $10\log_{10}(G)$ from ABSM protocol for a Type-II link, shown in comparison to the balancing loss $-10\log_{10}(\hat\pi)$ for the standard BSM protocol ($\sigma=1$).  The center of the plot corresponds to balanced losses between the sources and the BSMs.  The top left corresponds to both BSMs located near the central source $S_{34}$ and the bottom right corresponds to BSMs located close to the outer sources $S_{12}$ and $S_{56}$.  A maximum value of $\hat\pi=1.86$ is attained with the BSMs slightly biased towards the outer sources due to the higher Bell state fidelity of the (1,2,1)-error.} 
\label{fig:absmgain}
\end{figure}

The efficiency is further reduced in the presence of imbalanced losses between the sources and BSMs.  The additional loss factor $\hat\pi$ is shown in Figure \ref{fig:absmgain} for a link with 40 dB combined losses $\eta_2\eta_3=\eta_4\eta_5=0.01$.  The result is an additional 20 to 30 dB loss for the fully imbalanced link, i.e., the compensation required to balance the link reduces the combined source efficiency by a factor on the order of the channel efficiency ratios $\eta_2/\eta_3$ and $\eta_4/\eta_5$.

To understand the gain in source efficiency provided by the ABSM protocol we return to the expression \eqref{eq:maxeff} and find that the gain can be written
\small
\begin{equation}
G = \frac{bw(b)[36+27w(b)]^3\vert_{\sigma=1}}{bw(b)[36+27w(b)]^3\vert_{\sigma=0}}\Big(1 + \frac{1}{\bij_{23} + \bij_{54} + \bij_{23}\bij_{54}}\Big)
\end{equation}
\normalsize
which depends only on the parameters 
\begin{equation}
\bij_{ij}=\frac{\eta_i(1-\eta_j)}{\eta_j(1-\eta_i)}
\end{equation} 
characterizing the difference in channel losses adjacent to each BSM.  This dependence is made more explicit by expressing the parameter $b$ in the form
\begin{equation}
b = \frac{\sigma +\bij_{23} + \bij_{54} + \bij_{23}\bij_{54}}{9+3\bij_{23} + 3\bij_{54} + \bij_{23}\bij_{54}}.
\end{equation}
The result is shown in Figure \ref{fig:absmgain} for the same link with 40 dB combined channel loss.  As expected, the ABSM protocol significantly mitigates the balancing loss when both BSMs are biased towards the central source $\bij_{23},\bij_{54}<1$.

If both links are balanced, the ABSM protocol yields a 1.4 dB gain, with an increase to 3 dB if only one BSM is closer to the central source.  The general behavior can be summarized by the limiting cases
\begin{equation}
    G \simeq \begin{cases}  1.1\Big(1 + \dfrac{1}{2\bij_{23}}\Big), & \bij_{23}=\bij_{54} \ll 1 \\
    1, & \bij_{23}=\bij_{54} \gg 1,
    \\
     2.1, \quad\quad\;\;\;\;  & \bij_{23} \ll \bij_{54} = 1 \\
    1.4, \quad\quad\;\;\;\;  & \bij_{23} = \bij_{54} = 1 \\
    1, & \bij_{23} \gg \bij_{54} = 1.
    \end{cases}
\end{equation}

Unfortunately, the ABSM protocol does not yield any gain for the type of imbalance inherent in the satellite-assisted downlink $(\lambda_{23},\lambda_{54}\gg 1)$.  This is to be expected, since the primary double-pair errors in this case arise from double-pairs produced by the outer sources which go unchecked by the ABSM protocol.  However, by employing cascaded PDC sources in place of $S_{12}$ and $S_{56}$---which themselves perform an internal BSM---the ABSM protocol can be used to reduce the double-pair errors associated to the imbalance $\lambda_{23},\lambda_{54}\gg 1$.

The Type-II link with cascaded PDC sources is analyzed in the next section.  Before proceeding to that analysis, we note that another technique for achieving a Type-II link commensurate with a Type-I link is to equip the receivers at $A$ and $B$ with QND measurements with PNR to discard multi-photon events from sources $S_{12}$ and $S_{56}$, respectively.  The maximum efficiency $\hat\eta_{AB}$ then increases with the higher emission probabilities $p_{12}$ and $p_{56}$ afforded by the ability to identify and discard double-pair emission events from the end nodes.  The effect of receiver PNR is to reduce $\beta^{(2,1,1)}$, $\beta^{(1,1,2)}$, and $\beta^{(2,0,2)}$---describing the probability of a nominally successful double-swap given the corresponding photon numbers from each source---by a corresponding constant factor.  Specifically, $\beta^{(2,1,1)}$ and $\beta^{(1,1,2)}$ are reduced by the factor $(1-\alpha)$, where $\alpha$ represents the proportion of two-photon emission events from the adjacent source which are identified by the receiver, while $\beta^{(2,0,2)}$ is reduced by $(1-\alpha)^2$.  The parameter $b$ defined by \eqref{eq:bparameter} is left invariant and hence the maximum efficiency \eqref{eq:maxeff} is simply increased by $(1-\alpha)^{-2}$ independent of channel losses $\eta_i$ and BSM protocol $\sigma$.

\subsection{ABSM gain for a Type-II elementary link with cascaded PDC sources}

In the previous section, it was shown that the ABSM protocol provides the most significant gain for double-swap links where the BSMs are close to the central source.  As noted above, this is contrary to the type of imbalance inherent in the Type-II elementary link motivated by a satellite downlink.  Nevertheless, we now show that by introducing an additional BSM at each end of the link one can take advantage of the ABSM protocol to enable a Type-II downlink.

\begin{figure}[h!]
\centering\includegraphics[width=8.7cm]{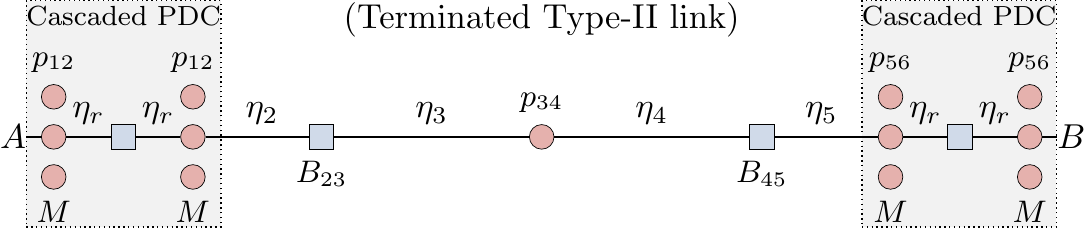}
\captionsetup{font=footnotesize,labelfont=footnotesize,justification=raggedright}
\caption{A terminated Type-II link with cascaded sources motivated by satellite-assisted entanglement distribution.  To maximize the utilization of photons from a satellite downlink, multiplexed cascaded PDC sources first use an internal BSM to herald production of an entangled pair before interfering with a photon from a satellite downlink.}
\label{fig:cascadedtype2}
\end{figure}

The basic idea is to employ a multiplexed, cascaded PDC source at both ends of the link to suppress double-pair emissions from the end nodes (Fig. \ref{fig:cascadedtype2}).  

The principle of a cascaded source is to use an active optical switch to route entangled pairs produced by a bank of PDC sources into a single output based on a successful BSM performed internally within the multiplexed source \cite{DHARA2022}.  This can be modeled as an inverted repeater architecture which is equivalent to the Type-II elementary link shown in Figure \ref{fig:linktypes} except that the repeater nodes now represent cascaded sources, with the `repeater node' swaps performed \emph{before} the BSMs interfering with the downlink.  

To model the efficiency of this architecture, we shall assume that the cascaded sources $S_{12}$ and $S_{56}$ are configured such that all internal sources have the same emission probability $p_{12}$ and $p_{56}$, respectively.  Following the procedure of the previous section we find the optimization follows an essentially equivalent mathematical form, with the optimal probabilities satisfying the same relation \eqref{eq:relativepp} with a redefinition of the weights given by (see SM)
\small
\begin{align}
b_{12} &= \frac{1}{2}\big[ \beta^{(21111)} + \frac{3}{4}\beta^{(12111)} - \frac{1}{2}\hat\beta^{(12111)} \big] 
\\
b_{56} &= \frac{1}{2}\big[ \beta^{(11112)} + \frac{3}{4}\beta^{(11121)} - \frac{1}{2}\hat\beta^{(11121)} \big] 
\\
b_{34} &= \frac{3}{4}\big[ \beta^{(11211)} - \frac{2}{3}\hat\beta^{(11211)} + \beta^{(20211)} \\
&\qquad\qquad + \beta^{(11202)} + \frac{3}{4}\beta^{(20202)}  \big] 
\end{align}
\normalsize
and
\begin{equation}
b = \frac{9}{16}\frac{b_{34}\beta^{(12021)}}{b_{12}b_{56}}
\end{equation}
determining the balancing of the multi-photon noise.  The parameter $p$ defined by \eqref{eq:relativepp} takes the modified form
\begin{equation}
p = \frac{8\beta^{(11111)}}{30 + 15bw(b)} \Delta f + O\big((\Delta f)^2 \big).
\end{equation}
The resulting maximum efficiency can again be factored in the form of \eqref{eq:factoredeff} as
\small
\begin{equation}
\hat\eta_{AB}(\Delta f) = \pi_0 \frac{1}{4}\eta_2\eta_3\eta_4\eta_5 \hat\pi(\eta_2,\eta_3,\eta_4,\eta_5) + O\big((\Delta f)^6\big).
\end{equation} 
\normalsize
The source efficiency $\pi_0$ in the low-efficiency limit is
\begin{equation}
\pi_0 = \frac{M^2\eta_r^4}{(1-\eta_r)^4} (\Delta f)^5 \times\begin{cases}
3\cdot 10^{-8}, & \sigma = 1, \\
2\cdot 10^{-7}, & \sigma = 0,
\end{cases}
\end{equation}
where $M$ is the number of pairs of PDC sources multiplexed in each cascaded source and $\eta_r$ is the internal efficiency.  To simplify the analysis we assume $M$ is not too large relative to the efficiency of the cascaded source internal swap $M\ll 32/(\eta_r\Delta f)^2$.  The benefit of introducing the cascaded sources is to give some engineering control over the Type-II link efficiency, obtained via the multiplexing factor $M$ and internal efficiency $\eta_r$ of the cascaded source.  
It should be observed that a high degree of multiplexing is required to overcome the $10^{-5}(\Delta f)^2$ efficiency reduction relative to the passive double swap link \eqref{eq:pi0doubleswap} or the $4\cdot 10^{-6}(\Delta f)^3$ reduction relative to the Type-I single-swap link.  As the multiplexing factor approaches the internal swap efficiency $M\sim 32/(\eta_r\Delta f)^2$, the cascaded source becomes a nearly deterministic source and a more detailed analysis is required \cite{DHARA2022}.

\begin{figure}[h!]
\centering\includegraphics[width=8.0cm]{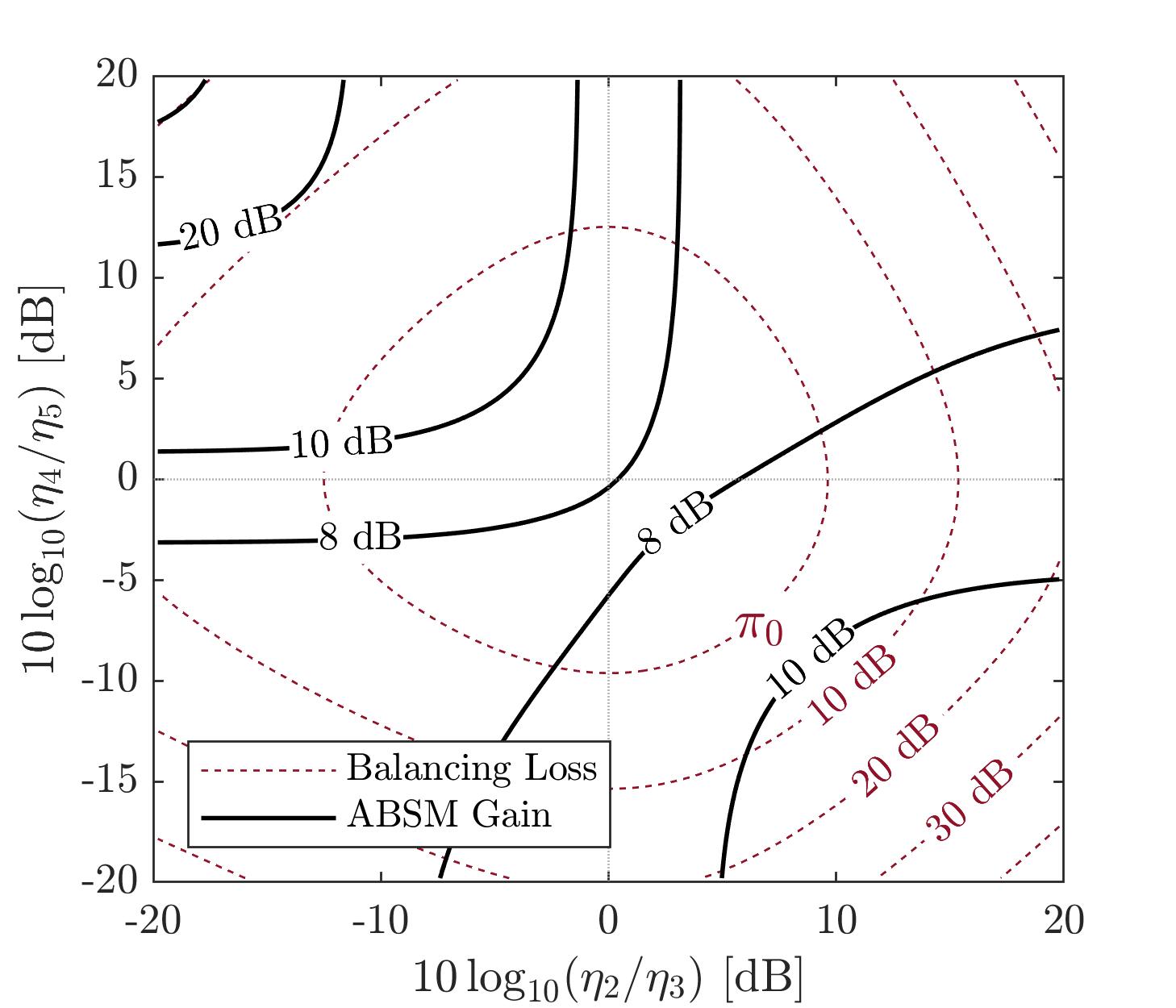}
\captionsetup{font=footnotesize,labelfont=footnotesize,justification=raggedright}
\caption{Gain in link efficiency $10\log_{10}(G)$ from ABSM protocol for a Type-II link with cascaded PDC sources.  The result is shown against the balancing loss $-10\log_{10}(\hat\pi)$ for the non-ABSM protocol ($\sigma=1$).  The center of the plot corresponds to balanced losses between the sources and the BSMs where a maximum value of $\hat\pi=6.3$ is attained.} 
\label{fig:absmgaincascaded}
\end{figure}

For a Type-II link with the BSMs located close to the cascaded sources, the balancing loss must also be taken into account and is shown in Figure \ref{fig:absmgaincascaded} for the same 40 dB combined loss as in Figure \ref{fig:absmgain}.  Figure \ref{fig:absmgaincascaded} also shows the gain provided by the ABSM protocol, in direct correspondence to the Type-II link shown in Figure \ref{fig:absmgain}.  The presence of the `repeater nodes' (i.e. cascaded sources) at the ends of the Type-II link allows the ABSM protocol to be employed to discard multi-photon events in the two central BSMs.  This compensates the balancing loss with up to 12 dB gain for BSMs close to the cascaded sources.

As an example, for a fully imbalanced Type-II link with 40 dB combined loss targeting a 95\% Bell state fidelity, the passive double-swap using PDC sources analyzed in the previous section yields a 48 dB efficiency reduction relative to a balanced Type-I link with the same combined loss.  With a multiplexing factor $M=1000$ and internal efficiency $\eta_r=0.95$ the same Type-II link with cascaded sources reduces this to 31 dB if the standard BSM protocol is used.  This is further reduced to 19 dB if the ABSM protocol is employed, enabling a Type-II link which is potentially comparable to a standard Type-I elementary link when the latter has implementation constraints as in a satellite-assisted entanglement distribution application.

\section{Conclusion}

In the absence of high-rate, environmentally robust, deterministic sources of entangled photon pairs, sources based on PDC remain one of the most promising candidates for high-rate entanglement distribution, particularly for satellite-assisted links.  However, multi-photon noise presents a significant limitation to the performance of entanglement swapping links based on PDC.  Previous attempts to address this issue are based on the use of PNR detection to identify and discard multi-photon emissions \cite{GUHA2015,DHARA2022}.  By taking advantage of quantum correlations present in the 4-photon term of the entangled TMSV state, the ABSM protocol introduced in this paper represents the only technique of which the authors are aware which suppresses double-pair noise from PDC-based entangled pair sources which does not rely on PNR detection.  Furthermore, the protocol can be employed in conjunction with other techniques such as cascaded PDC sources to improve the double-pair suppression afforded by PNR (Section \ref{sec:absmgain}).  The multi-photon noise suppression provided by the protocol analyzed in this work applies to any links using entangled pair sources with stimulated multi-photon emissions represented by a TMSV state.

Specifically, the ABSM protocol was shown to eliminate the dominant quadratic growth of multi-photon errors with the number of elementary links $\ell$ in repeater chains with PDC sources, yielding a gain in elementary link efficiency which grows quadratically with the number of elementary links in the chain \eqref{eq:bsfidelitychain}-\eqref{eq:repeatergain}.  The analysis also introduced a new calculus for evaluating key observables of the quantum state produced by repeater chains of arbitrary length (Section \ref{sec:linkmodel}).  In particular, a compact, closed form expression for the Bell state fidelity was presented including multi-photon terms \eqref{eq:bsfidelitychain}.  This calculus was also employed to obtain analytical calculations of the optimal emission probabilities and maximum efficiency for concatenated entanglement swapping links with imbalanced channel losses. 

Finally, it was shown that the ABSM protocol yields a 12 dB gain for Type-II elementary entanglement distribution links with cascaded PDC sources.  Combined with the multi-pair suppression provided by PNR in the cascaded sources, it was shown that with sufficient multiplexing, such sources can enable a Type-II elementary link with performance comparable to a standard Type-I link.  Although at present the largest multiplexed PDC sources that have been demonstrated consist of no more than ${\sim}$10 spatially multiplexed sources \cite{SCOTT2020}, such highly multiplexed sources constitute an engineering task for ground-based technology which can reduce flight system complexity for satellite-assisted entanglement distribution.  

\section*{Acknowledgements}

The authors would like to thank Evan Katz and Adam Fallon for a number of useful discussions during the preparation of this work.  This research was supported by the NASA Space Communications and Navigation (SCaN) Program and the NASA Glenn Research Center Communications \& Intelligent Systems Division.
\vfill

\bibliography{ChahineBibliography2022}

\begin{thebibliography}{20}%
\makeatletter
\providecommand \@ifxundefined [1]{%
 \@ifx{#1\undefined}
}%
\providecommand \@ifnum [1]{%
 \ifnum #1\expandafter \@firstoftwo
 \else \expandafter \@secondoftwo
 \fi
}%
\providecommand \@ifx [1]{%
 \ifx #1\expandafter \@firstoftwo
 \else \expandafter \@secondoftwo
 \fi
}%
\providecommand \natexlab [1]{#1}%
\providecommand \enquote  [1]{``#1''}%
\providecommand \bibnamefont  [1]{#1}%
\providecommand \bibfnamefont [1]{#1}%
\providecommand \citenamefont [1]{#1}%
\providecommand \href@noop [0]{\@secondoftwo}%
\providecommand \href [0]{\begingroup \@sanitize@url \@href}%
\providecommand \@href[1]{\@@startlink{#1}\@@href}%
\providecommand \@@href[1]{\endgroup#1\@@endlink}%
\providecommand \@sanitize@url [0]{\catcode `\\12\catcode `\$12\catcode
  `\&12\catcode `\#12\catcode `\^12\catcode `\_12\catcode `\%12\relax}%
\providecommand \@@startlink[1]{}%
\providecommand \@@endlink[0]{}%
\providecommand \url  [0]{\begingroup\@sanitize@url \@url }%
\providecommand \@url [1]{\endgroup\@href {#1}{\urlprefix }}%
\providecommand \urlprefix  [0]{URL }%
\providecommand \Eprint [0]{\href }%
\providecommand \doibase [0]{https://doi.org/}%
\providecommand \selectlanguage [0]{\@gobble}%
\providecommand \bibinfo  [0]{\@secondoftwo}%
\providecommand \bibfield  [0]{\@secondoftwo}%
\providecommand \translation [1]{[#1]}%
\providecommand \BibitemOpen [0]{}%
\providecommand \bibitemStop [0]{}%
\providecommand \bibitemNoStop [0]{.\EOS\space}%
\providecommand \EOS [0]{\spacefactor3000\relax}%
\providecommand \BibitemShut  [1]{\csname bibitem#1\endcsname}%
\let\auto@bib@innerbib\@empty
\bibitem [{\citenamefont {Kómár}\ \emph {et~al.}(2014)\citenamefont
  {Kómár}, \citenamefont {Kessler}, \citenamefont {Bishof}, \citenamefont
  {Jiang}, \citenamefont {Sørensen}, \citenamefont {Ye},\ and\ \citenamefont
  {Lukin}}]{KOMAR2014}%
  \BibitemOpen
  \bibfield  {author} {\bibinfo {author} {\bibfnamefont {P.}~\bibnamefont
  {Kómár}}, \bibinfo {author} {\bibfnamefont {E.~M.}\ \bibnamefont
  {Kessler}}, \bibinfo {author} {\bibfnamefont {M.}~\bibnamefont {Bishof}},
  \bibinfo {author} {\bibfnamefont {L.}~\bibnamefont {Jiang}}, \bibinfo
  {author} {\bibfnamefont {A.~S.}\ \bibnamefont {Sørensen}}, \bibinfo {author}
  {\bibfnamefont {J.}~\bibnamefont {Ye}},\ and\ \bibinfo {author}
  {\bibfnamefont {M.~D.}\ \bibnamefont {Lukin}},\ }\bibfield  {title} {\bibinfo
  {title} {A quantum network of clocks},\ }\href
  {https://doi.org/10.1038/nphys3000} {\bibfield  {journal} {\bibinfo
  {journal} {Nature Physics}\ }\textbf {\bibinfo {volume} {10}},\ \bibinfo
  {pages} {582} (\bibinfo {year} {2014})}\BibitemShut {NoStop}%
\bibitem [{\citenamefont {Nichol}\ \emph {et~al.}(2022)\citenamefont {Nichol},
  \citenamefont {Srinivas}, \citenamefont {Nadlinger}, \citenamefont {Drmota},
  \citenamefont {Main}, \citenamefont {Araneda}, \citenamefont {Ballance},\
  and\ \citenamefont {Lucas}}]{NICHOL2022}%
  \BibitemOpen
  \bibfield  {author} {\bibinfo {author} {\bibfnamefont {B.~C.}\ \bibnamefont
  {Nichol}}, \bibinfo {author} {\bibfnamefont {R.}~\bibnamefont {Srinivas}},
  \bibinfo {author} {\bibfnamefont {D.~P.}\ \bibnamefont {Nadlinger}}, \bibinfo
  {author} {\bibfnamefont {P.}~\bibnamefont {Drmota}}, \bibinfo {author}
  {\bibfnamefont {D.}~\bibnamefont {Main}}, \bibinfo {author} {\bibfnamefont
  {G.}~\bibnamefont {Araneda}}, \bibinfo {author} {\bibfnamefont {C.~J.}\
  \bibnamefont {Ballance}},\ and\ \bibinfo {author} {\bibfnamefont {D.~M.}\
  \bibnamefont {Lucas}},\ }\bibfield  {title} {\bibinfo {title} {An elementary
  quantum network of entangled optical atomic clocks},\ }\href
  {https://doi.org/10.1038/s41586-022-05088-z} {\bibfield  {journal} {\bibinfo
  {journal} {Nature}\ }\textbf {\bibinfo {volume} {609}},\ \bibinfo {pages}
  {689} (\bibinfo {year} {2022})}\BibitemShut {NoStop}%
\bibitem [{\citenamefont {Zhang}\ and\ \citenamefont
  {Zhuang}(2021)}]{ZHANG2021}%
  \BibitemOpen
  \bibfield  {author} {\bibinfo {author} {\bibfnamefont {Z.}~\bibnamefont
  {Zhang}}\ and\ \bibinfo {author} {\bibfnamefont {Q.}~\bibnamefont {Zhuang}},\
  }\bibfield  {title} {\bibinfo {title} {Distributed quantum sensing},\ }\href
  {https://doi.org/10.1088/2058-9565/abd4c3} {\bibfield  {journal} {\bibinfo
  {journal} {Quantum Science and Technology}\ }\textbf {\bibinfo {volume}
  {6}},\ \bibinfo {pages} {043001} (\bibinfo {year} {2021})}\BibitemShut
  {NoStop}%
\bibitem [{\citenamefont {Gottesman}\ \emph {et~al.}(2012)\citenamefont
  {Gottesman}, \citenamefont {Jennewein},\ and\ \citenamefont
  {Croke}}]{GOTTESMAN2012}%
  \BibitemOpen
  \bibfield  {author} {\bibinfo {author} {\bibfnamefont {D.}~\bibnamefont
  {Gottesman}}, \bibinfo {author} {\bibfnamefont {T.}~\bibnamefont
  {Jennewein}},\ and\ \bibinfo {author} {\bibfnamefont {S.}~\bibnamefont
  {Croke}},\ }\bibfield  {title} {\bibinfo {title} {Longer-baseline telescopes
  using quantum repeaters},\ }\href
  {https://doi.org/10.1103/PhysRevLett.109.070503} {\bibfield  {journal}
  {\bibinfo  {journal} {Phys. Rev. Lett.}\ }\textbf {\bibinfo {volume} {109}},\
  \bibinfo {pages} {070503} (\bibinfo {year} {2012})}\BibitemShut {NoStop}%
\bibitem [{\citenamefont {Mohageg}\ \emph {et~al.}(2022)\citenamefont
  {Mohageg}, \citenamefont {Mazzarella}, \citenamefont {Anastopoulos},
  \citenamefont {Gallicchio}, \citenamefont {Hu}, \citenamefont {Jennewein},
  \citenamefont {Johnson}, \citenamefont {Lin}, \citenamefont {Ling},
  \citenamefont {Marquardt}, \citenamefont {Meister}, \citenamefont {Newell},
  \citenamefont {Roura}, \citenamefont {Schleich}, \citenamefont {Schubert},
  \citenamefont {Strekalov}, \citenamefont {Vallone}, \citenamefont
  {Villoresi}, \citenamefont {Wörner}, \citenamefont {Yu}, \citenamefont
  {Zhai},\ and\ \citenamefont {Kwiat}}]{MOHAGEG2022}%
  \BibitemOpen
  \bibfield  {author} {\bibinfo {author} {\bibfnamefont {M.}~\bibnamefont
  {Mohageg}}, \bibinfo {author} {\bibfnamefont {L.}~\bibnamefont {Mazzarella}},
  \bibinfo {author} {\bibfnamefont {C.}~\bibnamefont {Anastopoulos}}, \bibinfo
  {author} {\bibfnamefont {J.}~\bibnamefont {Gallicchio}}, \bibinfo {author}
  {\bibfnamefont {B.-L.}\ \bibnamefont {Hu}}, \bibinfo {author} {\bibfnamefont
  {T.}~\bibnamefont {Jennewein}}, \bibinfo {author} {\bibfnamefont
  {S.}~\bibnamefont {Johnson}}, \bibinfo {author} {\bibfnamefont {S.-Y.}\
  \bibnamefont {Lin}}, \bibinfo {author} {\bibfnamefont {A.}~\bibnamefont
  {Ling}}, \bibinfo {author} {\bibfnamefont {C.}~\bibnamefont {Marquardt}},
  \bibinfo {author} {\bibfnamefont {M.}~\bibnamefont {Meister}}, \bibinfo
  {author} {\bibfnamefont {R.}~\bibnamefont {Newell}}, \bibinfo {author}
  {\bibfnamefont {A.}~\bibnamefont {Roura}}, \bibinfo {author} {\bibfnamefont
  {W.~P.}\ \bibnamefont {Schleich}}, \bibinfo {author} {\bibfnamefont
  {C.}~\bibnamefont {Schubert}}, \bibinfo {author} {\bibfnamefont {D.~V.}\
  \bibnamefont {Strekalov}}, \bibinfo {author} {\bibfnamefont {G.}~\bibnamefont
  {Vallone}}, \bibinfo {author} {\bibfnamefont {P.}~\bibnamefont {Villoresi}},
  \bibinfo {author} {\bibfnamefont {L.}~\bibnamefont {Wörner}}, \bibinfo
  {author} {\bibfnamefont {N.}~\bibnamefont {Yu}}, \bibinfo {author}
  {\bibfnamefont {A.}~\bibnamefont {Zhai}},\ and\ \bibinfo {author}
  {\bibfnamefont {P.}~\bibnamefont {Kwiat}},\ }\bibfield  {title} {\bibinfo
  {title} {The deep space quantum link: prospective fundamental physics
  experiments using long-baseline quantum optics},\ }\bibfield  {journal}
  {\bibinfo  {journal} {EPJ Quantum Technology}\ }\textbf {\bibinfo {volume}
  {9}},\ \href {https://doi.org/10.1140/epjqt/s40507-022-00143-0}
  {10.1140/epjqt/s40507-022-00143-0} (\bibinfo {year} {2022})\BibitemShut
  {NoStop}%
\bibitem [{\citenamefont {Sinclair}\ \emph {et~al.}(2014)\citenamefont
  {Sinclair}, \citenamefont {Saglamyurek}, \citenamefont {Mallahzadeh},
  \citenamefont {Slater}, \citenamefont {George}, \citenamefont {Ricken},
  \citenamefont {Hedges}, \citenamefont {Oblak}, \citenamefont {Simon},
  \citenamefont {Sohler},\ and\ \citenamefont {Tittel}}]{SINCLAIR2010}%
  \BibitemOpen
  \bibfield  {author} {\bibinfo {author} {\bibfnamefont {N.}~\bibnamefont
  {Sinclair}}, \bibinfo {author} {\bibfnamefont {E.}~\bibnamefont
  {Saglamyurek}}, \bibinfo {author} {\bibfnamefont {H.}~\bibnamefont
  {Mallahzadeh}}, \bibinfo {author} {\bibfnamefont {J.~A.}\ \bibnamefont
  {Slater}}, \bibinfo {author} {\bibfnamefont {M.}~\bibnamefont {George}},
  \bibinfo {author} {\bibfnamefont {R.}~\bibnamefont {Ricken}}, \bibinfo
  {author} {\bibfnamefont {M.~P.}\ \bibnamefont {Hedges}}, \bibinfo {author}
  {\bibfnamefont {D.}~\bibnamefont {Oblak}}, \bibinfo {author} {\bibfnamefont
  {C.}~\bibnamefont {Simon}}, \bibinfo {author} {\bibfnamefont
  {W.}~\bibnamefont {Sohler}},\ and\ \bibinfo {author} {\bibfnamefont
  {W.}~\bibnamefont {Tittel}},\ }\bibfield  {title} {\bibinfo {title} {Spectral
  multiplexing for scalable quantum photonics using an atomic frequency comb
  quantum memory and feed-forward control},\ }\href
  {https://doi.org/10.1103/PhysRevLett.113.053603} {\bibfield  {journal}
  {\bibinfo  {journal} {Phys. Rev. Lett.}\ }\textbf {\bibinfo {volume} {113}},\
  \bibinfo {pages} {053603} (\bibinfo {year} {2014})}\BibitemShut {NoStop}%
\bibitem [{\citenamefont {Sangouard}\ \emph {et~al.}(2011)\citenamefont
  {Sangouard}, \citenamefont {Simon}, \citenamefont {de~Riedmatten},\ and\
  \citenamefont {Gisin}}]{SANGOUARD2011}%
  \BibitemOpen
  \bibfield  {author} {\bibinfo {author} {\bibfnamefont {N.}~\bibnamefont
  {Sangouard}}, \bibinfo {author} {\bibfnamefont {C.}~\bibnamefont {Simon}},
  \bibinfo {author} {\bibfnamefont {H.}~\bibnamefont {de~Riedmatten}},\ and\
  \bibinfo {author} {\bibfnamefont {N.}~\bibnamefont {Gisin}},\ }\bibfield
  {title} {\bibinfo {title} {Quantum repeaters based on atomic ensembles and
  linear optics},\ }\href {https://doi.org/10.1103/RevModPhys.83.33} {\bibfield
   {journal} {\bibinfo  {journal} {Rev. Mod. Phys.}\ }\textbf {\bibinfo
  {volume} {83}},\ \bibinfo {pages} {33} (\bibinfo {year} {2011})}\BibitemShut
  {NoStop}%
\bibitem [{\citenamefont {Guha}\ \emph {et~al.}(2015)\citenamefont {Guha},
  \citenamefont {Krovi}, \citenamefont {Fuchs}, \citenamefont {Dutton},
  \citenamefont {Slater}, \citenamefont {Simon},\ and\ \citenamefont
  {Tittel}}]{GUHA2015}%
  \BibitemOpen
  \bibfield  {author} {\bibinfo {author} {\bibfnamefont {S.}~\bibnamefont
  {Guha}}, \bibinfo {author} {\bibfnamefont {H.}~\bibnamefont {Krovi}},
  \bibinfo {author} {\bibfnamefont {C.~A.}\ \bibnamefont {Fuchs}}, \bibinfo
  {author} {\bibfnamefont {Z.}~\bibnamefont {Dutton}}, \bibinfo {author}
  {\bibfnamefont {J.~A.}\ \bibnamefont {Slater}}, \bibinfo {author}
  {\bibfnamefont {C.}~\bibnamefont {Simon}},\ and\ \bibinfo {author}
  {\bibfnamefont {W.}~\bibnamefont {Tittel}},\ }\bibfield  {title} {\bibinfo
  {title} {Rate-loss analysis of an efficient quantum repeater architecture},\
  }\href {https://doi.org/10.1103/PhysRevA.92.022357} {\bibfield  {journal}
  {\bibinfo  {journal} {Phys. Rev. A}\ }\textbf {\bibinfo {volume} {92}},\
  \bibinfo {pages} {022357} (\bibinfo {year} {2015})}\BibitemShut {NoStop}%
\bibitem [{\citenamefont {Pant}\ \emph {et~al.}(2017)\citenamefont {Pant},
  \citenamefont {Krovi}, \citenamefont {Englund},\ and\ \citenamefont
  {Guha}}]{PANT2017}%
  \BibitemOpen
  \bibfield  {author} {\bibinfo {author} {\bibfnamefont {M.}~\bibnamefont
  {Pant}}, \bibinfo {author} {\bibfnamefont {H.}~\bibnamefont {Krovi}},
  \bibinfo {author} {\bibfnamefont {D.}~\bibnamefont {Englund}},\ and\ \bibinfo
  {author} {\bibfnamefont {S.}~\bibnamefont {Guha}},\ }\bibfield  {title}
  {\bibinfo {title} {Rate-distance tradeoff and resource costs for all-optical
  quantum repeaters},\ }\href {https://doi.org/10.1103/PhysRevA.95.012304}
  {\bibfield  {journal} {\bibinfo  {journal} {Phys. Rev. A}\ }\textbf {\bibinfo
  {volume} {95}},\ \bibinfo {pages} {012304} (\bibinfo {year}
  {2017})}\BibitemShut {NoStop}%
\bibitem [{\citenamefont {Krovi}\ \emph {et~al.}(2016)\citenamefont {Krovi},
  \citenamefont {Guha}, \citenamefont {Dutton}, \citenamefont {Slater},
  \citenamefont {Simon},\ and\ \citenamefont {Tittel}}]{KROVI2016}%
  \BibitemOpen
  \bibfield  {author} {\bibinfo {author} {\bibfnamefont {H.}~\bibnamefont
  {Krovi}}, \bibinfo {author} {\bibfnamefont {S.}~\bibnamefont {Guha}},
  \bibinfo {author} {\bibfnamefont {Z.}~\bibnamefont {Dutton}}, \bibinfo
  {author} {\bibfnamefont {J.~A.}\ \bibnamefont {Slater}}, \bibinfo {author}
  {\bibfnamefont {C.}~\bibnamefont {Simon}},\ and\ \bibinfo {author}
  {\bibfnamefont {W.}~\bibnamefont {Tittel}},\ }\bibfield  {title} {\bibinfo
  {title} {Practical quantum repeaters with parametric down-conversion
  sources},\ }\bibfield  {journal} {\bibinfo  {journal} {Applied Physics B}\
  }\textbf {\bibinfo {volume} {122}},\ \href
  {https://doi.org/10.1007/s00340-015-6297-4} {10.1007/s00340-015-6297-4}
  (\bibinfo {year} {2016})\BibitemShut {NoStop}%
\bibitem [{\citenamefont {Dhara}\ \emph {et~al.}(2022)\citenamefont {Dhara},
  \citenamefont {Johnson}, \citenamefont {Gagatsos}, \citenamefont {Kwiat},\
  and\ \citenamefont {Guha}}]{DHARA2022}%
  \BibitemOpen
  \bibfield  {author} {\bibinfo {author} {\bibfnamefont {P.}~\bibnamefont
  {Dhara}}, \bibinfo {author} {\bibfnamefont {S.~J.}\ \bibnamefont {Johnson}},
  \bibinfo {author} {\bibfnamefont {C.~N.}\ \bibnamefont {Gagatsos}}, \bibinfo
  {author} {\bibfnamefont {P.~G.}\ \bibnamefont {Kwiat}},\ and\ \bibinfo
  {author} {\bibfnamefont {S.}~\bibnamefont {Guha}},\ }\bibfield  {title}
  {\bibinfo {title} {Heralded multiplexed high-efficiency cascaded source of
  dual-rail entangled photon pairs using spontaneous parametric
  down-conversion},\ }\href {https://doi.org/10.1103/PhysRevApplied.17.034071}
  {\bibfield  {journal} {\bibinfo  {journal} {Phys. Rev. Applied}\ }\textbf
  {\bibinfo {volume} {17}},\ \bibinfo {pages} {034071} (\bibinfo {year}
  {2022})}\BibitemShut {NoStop}%
\bibitem [{\citenamefont {Sangouard}\ \emph {et~al.}(2008)\citenamefont
  {Sangouard}, \citenamefont {Simon}, \citenamefont {Zhao}, \citenamefont
  {Chen}, \citenamefont {de~Riedmatten}, \citenamefont {Pan},\ and\
  \citenamefont {Gisin}}]{SANGOUARD2008}%
  \BibitemOpen
  \bibfield  {author} {\bibinfo {author} {\bibfnamefont {N.}~\bibnamefont
  {Sangouard}}, \bibinfo {author} {\bibfnamefont {C.}~\bibnamefont {Simon}},
  \bibinfo {author} {\bibfnamefont {B.}~\bibnamefont {Zhao}}, \bibinfo {author}
  {\bibfnamefont {Y.-A.}\ \bibnamefont {Chen}}, \bibinfo {author}
  {\bibfnamefont {H.}~\bibnamefont {de~Riedmatten}}, \bibinfo {author}
  {\bibfnamefont {J.-W.}\ \bibnamefont {Pan}},\ and\ \bibinfo {author}
  {\bibfnamefont {N.}~\bibnamefont {Gisin}},\ }\bibfield  {title} {\bibinfo
  {title} {Robust and efficient quantum repeaters with atomic ensembles and
  linear optics},\ }\href {https://doi.org/10.1103/PhysRevA.77.062301}
  {\bibfield  {journal} {\bibinfo  {journal} {Phys. Rev. A}\ }\textbf {\bibinfo
  {volume} {77}},\ \bibinfo {pages} {062301} (\bibinfo {year}
  {2008})}\BibitemShut {NoStop}%
\bibitem [{\citenamefont {Yu}\ \emph {et~al.}(2020)\citenamefont {Yu},
  \citenamefont {Ma}, \citenamefont {Luo}, \citenamefont {Jing}, \citenamefont
  {Sun}, \citenamefont {Fang}, \citenamefont {Yang}, \citenamefont {Liu},
  \citenamefont {Zheng}, \citenamefont {Xie}, \citenamefont {Zhang},
  \citenamefont {You}, \citenamefont {Wang}, \citenamefont {Chen},
  \citenamefont {Zhang}, \citenamefont {Bao},\ and\ \citenamefont
  {Pan}}]{YU2020}%
  \BibitemOpen
  \bibfield  {author} {\bibinfo {author} {\bibfnamefont {Y.}~\bibnamefont
  {Yu}}, \bibinfo {author} {\bibfnamefont {F.}~\bibnamefont {Ma}}, \bibinfo
  {author} {\bibfnamefont {X.-Y.}\ \bibnamefont {Luo}}, \bibinfo {author}
  {\bibfnamefont {B.}~\bibnamefont {Jing}}, \bibinfo {author} {\bibfnamefont
  {P.-F.}\ \bibnamefont {Sun}}, \bibinfo {author} {\bibfnamefont {R.-Z.}\
  \bibnamefont {Fang}}, \bibinfo {author} {\bibfnamefont {C.-W.}\ \bibnamefont
  {Yang}}, \bibinfo {author} {\bibfnamefont {H.}~\bibnamefont {Liu}}, \bibinfo
  {author} {\bibfnamefont {M.-Y.}\ \bibnamefont {Zheng}}, \bibinfo {author}
  {\bibfnamefont {X.-P.}\ \bibnamefont {Xie}}, \bibinfo {author} {\bibfnamefont
  {W.-J.}\ \bibnamefont {Zhang}}, \bibinfo {author} {\bibfnamefont {L.-X.}\
  \bibnamefont {You}}, \bibinfo {author} {\bibfnamefont {Z.}~\bibnamefont
  {Wang}}, \bibinfo {author} {\bibfnamefont {T.-Y.}\ \bibnamefont {Chen}},
  \bibinfo {author} {\bibfnamefont {Q.}~\bibnamefont {Zhang}}, \bibinfo
  {author} {\bibfnamefont {X.-H.}\ \bibnamefont {Bao}},\ and\ \bibinfo {author}
  {\bibfnamefont {J.-W.}\ \bibnamefont {Pan}},\ }\bibfield  {title} {\bibinfo
  {title} {Entanglement of two quantum memories via fibres over dozens of
  kilometres},\ }\href {https://doi.org/10.1038/s41586-020-1976-7} {\bibfield
  {journal} {\bibinfo  {journal} {Nature}\ }\textbf {\bibinfo {volume} {578}},\
  \bibinfo {pages} {240} (\bibinfo {year} {2020})}\BibitemShut {NoStop}%
\bibitem [{\citenamefont {Takeoka}\ \emph {et~al.}(2015)\citenamefont
  {Takeoka}, \citenamefont {Jin},\ and\ \citenamefont {Sasaki}}]{TAKEOKA2015}%
  \BibitemOpen
  \bibfield  {author} {\bibinfo {author} {\bibfnamefont {M.}~\bibnamefont
  {Takeoka}}, \bibinfo {author} {\bibfnamefont {R.-B.}\ \bibnamefont {Jin}},\
  and\ \bibinfo {author} {\bibfnamefont {M.}~\bibnamefont {Sasaki}},\
  }\bibfield  {title} {\bibinfo {title} {Full analysis of multi-photon pair
  effects in spontaneous parametric down conversion based photonic quantum
  information processing},\ }\href
  {https://doi.org/10.1088/1367-2630/17/4/043030} {\bibfield  {journal}
  {\bibinfo  {journal} {New Journal of Physics}\ }\textbf {\bibinfo {volume}
  {17}},\ \bibinfo {pages} {043030} (\bibinfo {year} {2015})}\BibitemShut
  {NoStop}%
\bibitem [{\citenamefont {Khalique}\ \emph {et~al.}(2013)\citenamefont
  {Khalique}, \citenamefont {Tittel},\ and\ \citenamefont
  {Sanders}}]{KHALIQUE2013}%
  \BibitemOpen
  \bibfield  {author} {\bibinfo {author} {\bibfnamefont {A.}~\bibnamefont
  {Khalique}}, \bibinfo {author} {\bibfnamefont {W.}~\bibnamefont {Tittel}},\
  and\ \bibinfo {author} {\bibfnamefont {B.~C.}\ \bibnamefont {Sanders}},\
  }\bibfield  {title} {\bibinfo {title} {Practical long-distance quantum
  communication using concatenated entanglement swapping},\ }\href
  {https://doi.org/10.1103/PhysRevA.88.022336} {\bibfield  {journal} {\bibinfo
  {journal} {Phys. Rev. A}\ }\textbf {\bibinfo {volume} {88}},\ \bibinfo
  {pages} {022336} (\bibinfo {year} {2013})}\BibitemShut {NoStop}%
\bibitem [{\citenamefont {Kok}\ and\ \citenamefont
  {Braunstein}(2000)}]{KOK2000}%
  \BibitemOpen
  \bibfield  {author} {\bibinfo {author} {\bibfnamefont {P.}~\bibnamefont
  {Kok}}\ and\ \bibinfo {author} {\bibfnamefont {S.~L.}\ \bibnamefont
  {Braunstein}},\ }\bibfield  {title} {\bibinfo {title} {Postselected versus
  nonpostselected quantum teleportation using parametric down-conversion},\
  }\href {https://doi.org/10.1103/PhysRevA.61.042304} {\bibfield  {journal}
  {\bibinfo  {journal} {Phys. Rev. A}\ }\textbf {\bibinfo {volume} {61}},\
  \bibinfo {pages} {042304} (\bibinfo {year} {2000})}\BibitemShut {NoStop}%
\bibitem [{\citenamefont {Lamas-Linares}\ \emph {et~al.}(2001)\citenamefont
  {Lamas-Linares}, \citenamefont {Howell},\ and\ \citenamefont
  {Bouwmeester}}]{LAMASLINARES2001}%
  \BibitemOpen
  \bibfield  {author} {\bibinfo {author} {\bibfnamefont {A.}~\bibnamefont
  {Lamas-Linares}}, \bibinfo {author} {\bibfnamefont {J.~C.}\ \bibnamefont
  {Howell}},\ and\ \bibinfo {author} {\bibfnamefont {D.}~\bibnamefont
  {Bouwmeester}},\ }\bibfield  {title} {\bibinfo {title} {Stimulated emission
  of polarization-entangled photons},\ }\href
  {https://doi.org/10.1038/35091014} {\bibfield  {journal} {\bibinfo  {journal}
  {Nature}\ }\textbf {\bibinfo {volume} {412}},\ \bibinfo {pages} {887}
  (\bibinfo {year} {2001})}\BibitemShut {NoStop}%
\bibitem [{\citenamefont {Durkin}\ \emph {et~al.}(2002)\citenamefont {Durkin},
  \citenamefont {Simon},\ and\ \citenamefont {Bouwmeester}}]{DURKIN2002}%
  \BibitemOpen
  \bibfield  {author} {\bibinfo {author} {\bibfnamefont {G.~A.}\ \bibnamefont
  {Durkin}}, \bibinfo {author} {\bibfnamefont {C.}~\bibnamefont {Simon}},\ and\
  \bibinfo {author} {\bibfnamefont {D.}~\bibnamefont {Bouwmeester}},\
  }\bibfield  {title} {\bibinfo {title} {Multiphoton entanglement concentration
  and quantum cryptography},\ }\href
  {https://doi.org/10.1103/PhysRevLett.88.187902} {\bibfield  {journal}
  {\bibinfo  {journal} {Phys. Rev. Lett.}\ }\textbf {\bibinfo {volume} {88}},\
  \bibinfo {pages} {187902} (\bibinfo {year} {2002})}\BibitemShut {NoStop}%
\bibitem [{\citenamefont {Shchukin}\ and\ \citenamefont {van
  Loock}(2022)}]{SHCHUKIN2022}%
  \BibitemOpen
  \bibfield  {author} {\bibinfo {author} {\bibfnamefont {E.}~\bibnamefont
  {Shchukin}}\ and\ \bibinfo {author} {\bibfnamefont {P.}~\bibnamefont {van
  Loock}},\ }\bibfield  {title} {\bibinfo {title} {Optimal entanglement
  swapping in quantum repeaters},\ }\href
  {https://doi.org/10.1103/PhysRevLett.128.150502} {\bibfield  {journal}
  {\bibinfo  {journal} {Phys. Rev. Lett.}\ }\textbf {\bibinfo {volume} {128}},\
  \bibinfo {pages} {150502} (\bibinfo {year} {2022})}\BibitemShut {NoStop}%
\bibitem [{\citenamefont {Meyer-Scott}\ \emph {et~al.}(2020)\citenamefont
  {Meyer-Scott}, \citenamefont {Silberhorn},\ and\ \citenamefont
  {Migdall}}]{SCOTT2020}%
  \BibitemOpen
  \bibfield  {author} {\bibinfo {author} {\bibfnamefont {E.}~\bibnamefont
  {Meyer-Scott}}, \bibinfo {author} {\bibfnamefont {C.}~\bibnamefont
  {Silberhorn}},\ and\ \bibinfo {author} {\bibfnamefont {A.}~\bibnamefont
  {Migdall}},\ }\bibfield  {title} {\bibinfo {title} {Single-photon sources:
  Approaching the ideal through multiplexing},\ }\href
  {https://doi.org/10.1063/5.0003320} {\bibfield  {journal} {\bibinfo
  {journal} {Review of Scientific Instruments}\ }\textbf {\bibinfo {volume}
  {91}},\ \bibinfo {pages} {041101} (\bibinfo {year} {2020})},\ \Eprint
  {https://arxiv.org/abs/https://doi.org/10.1063/5.0003320}
  {https://doi.org/10.1063/5.0003320} \BibitemShut {NoStop}%
\end{thebibliography}%

\section*{Notice for copyrighted information}
This manuscript is a work of the United States Government authored as part of the official duties of employee(s) of the National Aeronautics and Space Administration.  No copyright is claimed in the United States under Title 17, U.S. Code.  All other rights are reserved by the United States Government. Any publisher accepting this manuscript for publication acknowledges that the United States Government retains a non-exclusive, irrevocable, worldwide license to prepare derivative works, publish, or reproduce the published form of this manuscript, or allow others to do so, for United States government purposes.

\section{Supplementary Material}

\subsection{Equivalence of uniform losses before and after beam-splitter}\label{app:lossequivalence}

In this appendix we show that the detection efficiency $\eta_d$ can be combined with the channel transmission efficiency in the entanglement swapping link model described in Section \ref{sec:linkmodel}.  

\begin{figure}[h!]
\centering\includegraphics[width=7.0cm]{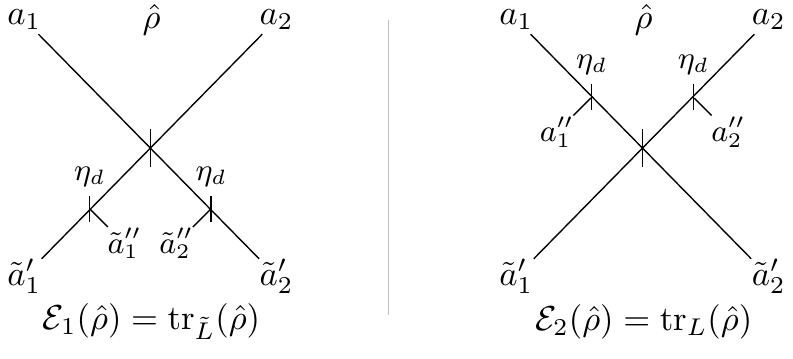}
\captionsetup{font=footnotesize,labelfont=footnotesize,justification=raggedright}
\caption{Two distinct physical systems which implement the same quantum operation $\mathcal E_1=\mathcal E_2$ on the state $\hat\rho$ of modes $\{a_1,a_2\}$ after taking the partial trace over the Hilbert space $\tilde L$ and $L$ generated by the loss modes $\{\tilde a_1'',\tilde a_2''\}$ and $\{a_1'',a_2''\}$, respectively.}
\label{fig:lossequivalence}
\end{figure}

Consider any state $\hat\rho$ consisting of bosonic excitations of two optical modes $a_1,a_2$ coupled to a pair of modes $\tilde a_1,\tilde a_2$ by a 50:50 beam-splitter
\begin{align}
a_1 &= \sqrt{1/2}(\tilde a_1 + \tilde a_2) \\
a_2 &= \sqrt{1/2}(\tilde a_1 - \tilde a_2).
\end{align}
We show the equivalence of two distinct physical systems (quantum operations on $\hat\rho$) shown in Figure \ref{fig:lossequivalence} in which uniform losses occur before or after the beam-splitter.

In the first system, we suppose that modes $\tilde a_1$ and $\tilde a_2$ are coupled to pure-loss bosonic channels with transmission efficiency $\eta_d$
\begin{align}
\tilde a_1 = \sqrt{\eta_d}\tilde a_1' + \sqrt{1-\eta_d}\tilde a_1'' \\
\tilde a_2 = \sqrt{\eta_d}\tilde a_2' + \sqrt{1-\eta_d}\tilde a_2''
\end{align}
while the modes $a_1,a_2$ are not subject to losses.  In the second system we suppose that instead the input modes $a_1,a_2$ are subject to pure-loss channels with efficiency $\eta_d$, while the output modes $\tilde a_1,\tilde a_2$ are not subject to losses.  We now show that the reduced state $\tr_{\tilde L}(\hat\rho)$ obtained by tracing over the Hilbert space $\tilde L$ generated by the loss modes $\{\tilde a_1'',\tilde a_2''\}$ of the first system is equivalent to the reduced state $\tr_{L}(\hat \rho)$ obtained by tracing over the Hilbert space $L$ generated by the loss modes $\{a_1'',a_2''\}$ of the second system.

To the first point, we may assume without loss of generality that $\hat\rho$ is a pure state of the form $|\psi\rangle = \mathcal P(a_1,a_2)|0\rangle$ where $\mathcal P$ is a polynomial in the operators $a_1,a_2$ with complex coefficients (appropriately normalized).  In the first system, applying a change of basis, the input modes take the form
\begin{align}\label{eq:a1s1}
a_1 &= \frac{1}{\sqrt{2}}( \sqrt{\eta_d}[\tilde a_1'+\tilde a_2']+\sqrt{1-\eta_d}[\tilde a_1'' + \tilde a_2''] ) \\
a_2 &= \frac{1}{\sqrt{2}}( \sqrt{\eta_d}[\tilde a_1'- \tilde a_2'] +\sqrt{1-\eta_d}[\tilde a_1'' - \tilde a_2''] )\label{eq:a2s1}
\end{align}
whereas in the second system, the input modes take the form
\begin{align}\label{eq:a1s2}
a_1 &= \frac{1}{\sqrt{2}}\sqrt{\eta_d}[\tilde a_1'+\tilde a_2']+\sqrt{1-\eta_d} a_1'' \\
a_2 &= \frac{1}{\sqrt{2}} \sqrt{\eta_d}[\tilde a_1'- \tilde a_2'] +\sqrt{1-\eta_d} a_2''.
\label{eq:a2s2}
\end{align}
Since the partial trace over $L,\tilde L$ may be taken in any orthonormal basis, we can apply the change of bases
\begin{align}
\hat a_1 = \sqrt{1/2}(\tilde a_1'' + \tilde a_2'') \\
\hat a_2 = \sqrt{1/2}(\tilde a_1'' - \tilde a_2'')
\end{align}
for the loss modes in the first system, taking $\hat a_1 = a_1''$ and $
\hat a_2 = a_2''$ in the second system, whereupon the input modes \eqref{eq:a1s1}-\eqref{eq:a2s1} become formally equivalent to the input modes \eqref{eq:a1s2}-\eqref{eq:a2s2}.  As a result, after taking the partial trace of $\hat\rho = \mathcal P(a_1,a_2)|0\rangle\langle 0|\mathcal P(a_1,a_2)^\dagger$ over the Hilbert space $L,\tilde L$ generated by the modes $\{\hat a_1,\hat a_2\}$ one obtains the same reduced state $\tr_L(\hat\rho) = \tr_{\tilde L}(\hat \rho)$.  Furthermore, the same proof directly generalizes to any state of the form $\mathcal P(a_1,a_2,b_1,b_2,...)|0\rangle$ where the state of the input modes $a_1,a_2$ may be correlated to the state of any number of orthogonal modes $b_i$.

\subsection{Bell state fidelity and BSM coincidence efficiency for concatenated entanglement swaps}

We now give a derivation of the expressions \eqref{eq:concateff}-\eqref{eq:concatefftrue} based on the model described in Section \ref{sec:linkmodel}.  The transmitted modes and loss modes from channel $i$ satisfy the relations given in the main text $a_i = \sqrt{\eta_i}a_i' + \sqrt{1-\eta_i}a_i''$ (similarly for $b_i$), and we denote the detection modes from the output of the beam splitter in the BSM $B_{ij}$ via the relations 
\begin{align}
a_i' = \sqrt{1/2}(\tilde a_i + \tilde a_j) \\
a_j' = \sqrt{1/2}(\tilde a_i - \tilde a_j) 
\end{align}
and similarly for the modes $b_i$.  As it is generally clear from context, we henceforth omit the prime on the transmitted modes and use $a_i,a_i'$ interchangeably.  Detection in each BSM detection mode (corresponding to a single optical mode $\tilde a_i,\tilde b_i,\tilde c_i,$ or $\tilde d_i$) is modeled via the POVM projectors
\begin{align}
D_0 &= \sum_{n=0}^4 P(0|n)|n\rangle\langle n|,\quad  D_1 = \sum_{n=0}^4 P(1|n)|n\rangle\langle n|
\end{align}
representing a detection of exactly $0$ or $1$ photon, respectively, in the corresponding optical mode.  Here, $P(m|n)$ represents the probability of detecting $m$ photons given an $n$-photon Fock state is present in the detection mode, and $|n\rangle\langle n |$ is the Fock state projector in the detection mode acting as the identity on all other modes.  Detectors with efficiency $\eta_d$ and PNR without extrinsic noise (background/dark counts) are modeled via
\begin{align}\label{eq:detectorpovm}
\begin{split}
P(0|n) &=  (1 - \eta_d)^n,\\
P(1|n) &= \sum_{k=0}^{n-1}\eta_d(1-\eta_d)^k(1 - \alpha\eta_d)^{n-1-k}
\end{split}
\end{align}
where we take $\alpha =1$ ($\alpha = 0$) to represent detection with (without) PNR.  In this work, we always assume PNR detection $\alpha=1$.  With $\alpha = 1$, this POVM is equivalent to a perfect detector $P(m|n)=\delta_{mn}$ situated behind a beam-splitter with transmission efficiency $\eta_d$, after tracing over the loss modes.  In accordance with the equivalence described in Appendix \ref{app:lossequivalence}, we henceforth assume perfect detectors $P(m|n)=\delta_{mn}$ and combine the detection efficiency with the channel efficiency $\eta_i\to \eta_i\eta_d$.

\subsubsection{Density matrix for reduced state after losses}\label{sec:reducedstate}

The reduced state $\hat \rho_{ij}$ for a single source $|\psi_{ij}\rangle$ obtained after tracing over all of the loss modes can be written
\begin{align}\label{eq:dmdecomp}
\begin{split}
\hat\rho_{ij} &= \rho_{1;1}^{(1)} + \rho_{1;1}^{(2)} + \rho_{0;0}^{(0)} + \rho_{0;0}^{(1)} + \rho_{0;0}^{(2)}  \\
&+ \rho_{1;0}^{(1)} + \rho_{1;0}^{(2)} + \rho_{0;1}^{(1)} + \rho_{0;1}^{(2)} \\
&+ \rho_{2;0}^{(2)} + \rho_{0;2}^{(2)} + \rho_{2;1}^{(2)} + \rho_{1;2}^{(2)} + \rho_{2;2}^{(2)}
\end{split}
\end{align}
where $\rho_{m;n}^{(\nu)}$ denotes the component of the density matrix which arises from an emission of $\nu$ photon pairs with losses resulting in an $m$-photon state in modes $a_i,b_i$ and an $n$ photon state in modes $a_j,b_j$.  In writing the density matrix in this form, we note that dephasing ensures that components with distinct total photon number $\nu$ add incoherently as density matrices (since we do not assume phase-stable channels, states of distinct energy become dephased after transmission), and for fixed $\nu$ the loss modes distinguish components of the state with distinct photon number $m;n$ in the output modes.  Throughout this work we shall adopt the convention that the hat notation $\hat\rho$ is reserved for normalized density matrices with unit trace.

We now calculate each component by expanding the state $|\psi_{ij}\rangle$ in the loss and output modes and grouping terms by the loss modes.  The main component is the entangled pair given by
\begin{equation}\label{eq:rho111}
\rho_{1;1}^{(1)} = p_{ij}^{(1)}\eta_i\eta_j |\Psi_{ij}^-\rangle_{ab}\langle \Psi_{ij}^-|_{ab}.
\end{equation}
The second component arises from two-pair emissions where one photon is lost from each output channel, and simplifies to the form
\small
\begin{equation}\label{eq:rho112}
\rho_{1;1}^{(2)} = 4p_{ij}^{(2)}\eta_i\eta_j(1-\eta_i)(1-\eta_j) \Big[ \frac{2}{3} |\Psi_{ij}^-\rangle\langle \Psi_{ij}^-| +\frac{1}{3} \hat\rho_{1;1}^{\textup{mix}}\Big]
\end{equation}
\normalsize
where $\hat\rho_{1;1}^{\textup{mix}}$ is the completely mixed two-photon state
\begin{align}
\begin{split}
\hat\rho_{1;1}^{\textup{mix}} &= \frac{1}{4}|a_ia_j\rangle \langle a_ia_j| + \frac{1}{4}|b_ib_j\rangle \langle b_ib_j| \\
&+ \frac{1}{4}|a_ib_j\rangle \langle a_ib_j| + \frac{1}{4}|b_ia_j\rangle \langle b_ia_j|.
\end{split}
\end{align}
and we use the notation $|a_i^n\rangle$ for the normalized state $\sqrt{1/n!}\;a_i^n |0\rangle$.  Note that given the state $\rho_{1;1}^{(2)}$, one obtains an entangled pair with probability $2/3$ and a completely mixed state with probability 1/3 which directly implies the result \eqref{eq:dpfidelity} for the fidelity of the state associated to tensor products of $\hat\rho_{1;1}^{(1)}$ and $\hat\rho_{1;1}^{(2)}$ (using linearity to expand the trace in the definition of the Bell state fidelity).

The vacuum terms are given by
\begin{align}
\rho_{0;0}^{(0)} &= p_{ij}^{(0)}|0\rangle\langle 0 |  \\
\rho_{0;0}^{(1)} &= p_{ij}^{(1)}(1-\eta_i)(1-\eta_j)|0\rangle\langle 0 | \\
\rho_{0;0}^{(2)} &= p_{ij}^{(2)}(1-\eta_i)^2(1-\eta_j)^2|0\rangle\langle 0 |.
\end{align}
For the remaining terms, we adopt a more compact notation for the density matrix as an ensemble of pure states, where we use the list $\{|\psi_1\rangle,|\psi_2\rangle,...,|\psi_m\rangle\}$ to denote the matrix $|\psi_1\rangle\langle \psi_1| + |\psi_2\rangle\langle\psi_2| + ...+|\psi_m\rangle\langle \psi_m|$.  The one-photon terms are given by
\begin{align}
\rho_{1;0}^{(1)} &= p_{ij}^{(1)}\eta_i(1-\eta_j)\Big[\frac{1}{2}\big\{|a_i\rangle,|b_i\rangle\big\}\Big] \\
\rho_{1;0}^{(2)} &= 2p_{ij}^{(2)}\eta_i(1-\eta_i)(1-\eta_j)^2\Big[\frac{1}{2} \big\{ |a_i\rangle, |b_i\rangle \big\}\Big]
\end{align}
where the trace has been factored outside the square brackets.  The terms $\rho_{m;n}^{(n)}$ are obtained from $\rho_{n;m}^{(n)}$ by the index substitution $i\leftrightarrow j$.  Keeping with the convention of factoring out the trace, the remaining terms
\begin{align}
&\rho_{2;0} = p_{ij}^{(2)} \eta_i^2(1-\eta_j)^2\Big[\frac{1}{3}\big\{ |a_i^2\rangle, |b_i^2\rangle, |a_ib_i\rangle \big\} \Big] \\
&\rho_{2;1} = 2p_{ij}^{(2)} \eta_i^2\eta_j(1-\eta_j) \\
&\;\Big[\frac{1}{6}\big\{ \sqrt{2} |a_i^2 b_j\rangle - |a_ib_ib_j\rangle, \sqrt{2} |b_i^2 a_j\rangle - |a_ib_ia_j\rangle \big\} \Big] \\
&\rho_{2;2} = p_{ij}^{(2)} \eta_i^2\eta_j^2\Big[\frac{1}{3}\big\{|a_i^2 b_j^2\rangle + |b_i^2a_j^2\rangle - |a_ib_ia_jb_j\rangle \big\} \Big]
\end{align}
complete the density matrix \eqref{eq:dmdecomp}.

In the next section, we shall find it convenient to decompose the density matrix according to the number of photon pairs emitted using \eqref{eq:dmdecomp}
\begin{equation}\label{eq:dmdecompnu}
\hat\rho_{ij} = \rho_{ij}^{(0)} + \rho_{ij}^{(1)} + \rho_{ij}^{(2)}
\end{equation}
with
\begin{equation}
\rho^{(\nu)}_{ij} = \sum_{m,n=0}^2 \rho_{m;n}^{(\nu)}.
\end{equation}

\subsubsection{Bell state measurement}

The outcome of the $k$-th BSM, adjacent to channels $(2k,2k+1)$, is modeled by a measurement POVM
\begin{equation}
\{ \tilde E_{k}^+, \tilde E_{k}^-, \tilde E_{k}^0\}
\end{equation}
where the operators $\tilde E_{k}^{\pm}$ are formed as a product of the operators $D_0$ and $D_1$ in the appropriate combinations of detection channels corresponding to a detection of the Bell states $|\Psi_{2k,2k+1}^\pm\rangle$, with $\tilde E_{k}^0=I - E_{k}^+ - E_{k}^-$ representing a failed BSM.  Two types of BSM operators are defined, distinguished by forming the constituent operators $\tilde E_{k}^{\pm}$ from detection of $|\Psi_{2k,2k+1}^\pm\rangle_{ab}$ or $|\Psi^{\pm}_{2k,2k+1}\rangle_{cd}$ in the mode basis $\{a,b\}$ or $\{c,d\}$, respectively.  Explicitly, for the $\{a,b\}$ basis we have
\begin{align}
\begin{split}\label{eq:eplus}
\tilde E_{k}^+ &= D_1(\tilde a_{2k})D_1(\tilde b_{2k})D_0(\tilde a_{2k+1})D_0(\tilde b_{2k+1}) \\
&+ D_0(\tilde a_{2k})D_0(\tilde b_{2k})D_1(\tilde a_{2k+1})D_1(\tilde b_{2k+1}),
\end{split}\\
\begin{split}
\tilde E_{k}^- &= D_1(\tilde a_{2k})D_0(\tilde b_{2k})D_0(\tilde a_{2k+1})D_1(\tilde b_{2k+1}) \\
&+ D_0(\tilde a_{2k})D_1(\tilde b_{2k})D_1(\tilde a_{2k+1})D_0(\tilde b_{2k+1}),
\end{split}\label{eq:eminus}
\end{align}
with the operators for the $\{c,d\}$ basis obtained by the substitution $a\leftrightarrow c$ and $b\leftrightarrow d$.

For a chain of successful BSMs, one can obtain a successful outcome from any chain of the operators $\tilde E_{k}^\pm$ for $k=1,...,N-1$.  The probability of a full chain of successful BSMs in the $\{a,b\}$ basis can thus be written
\begin{equation}\label{eq:tracebsm}
\bar\eta_{1,2N} = \tr(\tilde E\hat\rho)
\end{equation}
where $\hat\rho = \hat\rho_{12}\otimes \hat\rho_{34}\otimes ... \otimes \hat\rho_{2N-1,2N}$ and the operator
\begin{equation}\label{eq:bare}
\tilde E = \prod_{k=1}^{N-1} (E_{k}^+ + E_{k}^-)
\end{equation}
accounts for all possible sequences of successful BSMs.  

To make this trace analytically tractable, we make use of several properties.  First, we expand $\hat\rho_{ij}$ into a sum of tensor products of states $\rho_{ij}^{(\nu)}$ using the expansion \eqref{eq:dmdecompnu}.  Using linearity of the trace, we can then write \eqref{eq:tracebsm} as a sum of terms indexed by the vector $\vnu=(\nu_1,...,\nu_N)$ specifying the number of photon pairs emitted by each source.  Specifically, we define
\begin{equation}
\rho^{(\vnu)} = \bigotimes_{k=1}^N \rho_{2k-1,2k}^{(\nu_k)}
\end{equation}
and note that $\hat\rho = \sum_{\vnu} \rho^{(\vnu)}$.
  Second, since we can assume perfect detectors---recall the detection losses can be combined with the channel losses and we neglect extrinsic noise---the operator $\tilde E$ simply amounts to projecting out all terms except those with exactly two, oppositely polarized photons in each pair of BSM channels $(2k,2k+1)$.  This observation corresponds to the identity $\tilde E = E$ where $E$ is given by the same expressions \eqref{eq:eplus}, \eqref{eq:eminus}, and \eqref{eq:bare} except written in the source modes $\{a,b\}$ instead of the detection modes $\{\tilde a,\tilde b\}$.  This is verified rigorously by the identity
\small
\begin{align}
&\tilde E_{k}^+ + \tilde E_{k}^- = E_{k}^+ + E_{k}^- = \\
  &\quad\quad |a_{2k}b_{2k}\rangle\langle a_{2k}b_{2k}| + |a_{2k+1}b_{2k+1}\rangle\langle a_{2k+1}b_{2k+1}| \\
&\quad +  |a_{2k}b_{2k+1}\rangle\langle a_{2k}b_{2k+1}| + |a_{2k+1}b_{2k}\rangle\langle a_{2k+1}b_{2k}|
\end{align}
\normalsize
which expresses the fact that by counting successful BSM chains without distinguishing which Bell state was produced, we can perform the calculation with the source modes instead of the detection modes.

Working entirely in the basis of source modes $a_i,b_i$ with which we have already expressed the reduced state in the previous section, we note that the trace $\tr(E\hat\rho)$ now depends only on the density matrix elements on the main diagonal.  In this basis, the operator $E_{k}^+$ corresponds to a measurement of two oppositely polarized photons from a single source, and $E_{k}^-$ corresponds to a measurement of two oppositely polarized photons with one from each source.  This enables a convenient method for accounting for correlations between neighboring BSMs by introducing a formal operator
\begin{equation}\label{eq:be}
\mathbf E = \prod_{k=1}^{N} (\mathbf E_{k}^+ + \mathbf E_{k}^-)
\end{equation}
with
\begin{align}\label{eq:beplus}
\begin{split}
\mathbf  E_{k}^+ &= \mathbf e_{k}^{1}|a_{2k}b_{2k}\rangle\langle a_{2k}b_{2k}| + \mathbf e_{k}^{2}|a_{2k+1}b_{2k+1}\rangle\langle a_{2k+1}b_{2k+1}| \\
\mathbf  E_{k}^- &= \mathbf e_{k}^{3} |a_{2k}b_{2k+1}\rangle\langle a_{2k}b_{2k+1}| + \mathbf e_{k}^{4}|a_{2k+1}b_{2k}\rangle\langle a_{2k+1}b_{2k}|
\end{split}
\end{align}
where $\mathbf e_{k}^{m}$ simply represent indeterminates with $E_{k}^\pm$ obtained by setting $\mathbf e_{k}^{m}=1$.

The indeterminates $\mathbf e_{k}^{m}$ record the individual outcomes of the projection $E_{k}^\pm$ so that the trace can be expanded as
\small
\begin{align}
\begin{split}\label{eq:formaltrace}
\tr(\mathbf E \rho^{(\vnu)}) &= p^{(\vnu)}\sum_{\vec m} P_{\vnu}(\mathbf e_{1}^{m_1}\wedge...\wedge \mathbf e_{N-1}^{m_{N-1}}) \mathbf e_{1}^{m_1}\cdots\mathbf e_{N-1}^{m_{N-1}}
\end{split}
\end{align}
\normalsize
where $p^{(\vnu)}$ represents the probability that each source emits $\nu_i$ photon pairs---given by \eqref{eq:pprod} in the main text---and $P_{\vnu}$ denotes the conditional probability of the outcomes $\mathbf e_{k}^{m}$ given an emission of $\vnu$ photon pairs from each source, with the sum taken over all outcomes of the measurement $E$ at the input to the $N-1$ BSMs specified by $\vec m=(m_1,m_2,...,m_{N-1})$ with $1\leq m_k\leq 4$.  Having replaced the trace $\tr(\tilde E\hat\rho)$ by an equivalent trace $\tr(E\hat\rho)$ associated to a measurement POVM $E$ applied to modes at the input to the BSMs---where the state $\rho$ can be taken to be a diagonal matrix---we can now treat the correlations classically to write
\small
\begin{align}
\begin{split}\label{eq:psimultaneous}
&P_{\vnu}(\mathbf e_{1}^{m_1}\wedge...\wedge \mathbf e_{N-1}^{m_{N-1}})= \prod_{k=1}^{N-1} P_{\vnu}(\mathbf e_{k}^{m_k}) \prod_{j=2}^{N-1}\frac{ P_{\vnu}(\mathbf e_{j}^{m_j} | \mathbf e_{j-1}^{m_{j-1}})}{P_{\vnu}(\mathbf e_{j}^{m_j})}
\end{split}
\end{align}
\normalsize
where we use the fact that the outcome $\mathbf e_{k}^{m}$ is independent of the other outcomes with the exception of the adjacent BSMs $\mathbf e_{k-1}^{n}$ and $\mathbf e_{k+1}^{n}$.
Substituting \eqref{eq:psimultaneous} back into \eqref{eq:formaltrace}, the indeterminates $\mathbf e_{k}^{m}$ have served their original purpose of accounting all of the possible outcomes and we now repurpose them to write
\begin{equation}\label{eq:formaltrace2}
\tr(E\rho^{(\vnu)}) = p^{(\vnu)}\sum_{\vec m}\prod_{k=1}^{N-1} P_{\vnu}(\mathbf e_{k}^{m_k})\mathbf e_{k}^{m_k}
\end{equation}
where we return to the original operator $E$ and we have redefined the indeterminates $\mathbf e_{k}^{m}$ to satisfy the identity
\begin{equation}\label{eq:ekrelation}
\prod_{k=1}^{N-1} \mathbf e_{k}^{m_i} = \prod_{j=2}^{N-1}\frac{ P_{\vnu}(\mathbf e_{j}^{m_j} | \mathbf e_{j-1}^{m_{j-1}})}{P_{\vnu}(\mathbf e_{j}^{m_j})}.
\end{equation}
This representation allows us to formally factor the trace \eqref{eq:formaltrace2} as a product of the form
\begin{equation}\label{eq:factoredformaltrace}
\tr(E\rho^{(\vnu)}) = p^{(\vnu)}\prod_{k=1}^{N-1} \sum_{m=1}^4 P_{\vnu}(\mathbf e_{k}^{m})\mathbf e_{k}^{m} 
\end{equation}
where at present this factorization should be understood formally as it only has meaning when the product is expanded so that the substitution \eqref{eq:ekrelation} can be employed.

To make use of this factorization, we use the reduced state obtained in Section \ref{sec:reducedstate} to obtain the identities
\begin{equation}\label{eq:pekrelation1}
P_{\vnu}(\mathbf e_{k}^{3}) = P_{\vnu}(\mathbf e_{k}^{4})
\end{equation}
and for $k\geq 2$, either $P_{\vnu}(\mathbf e_{k}^{m})=0$ or
\small
\begin{equation}\hspace{-5pt}
\frac{P_{\vnu}(\mathbf e_{k}^{m} | \mathbf e_{k-1}^{n})}{P_{\vnu}(\mathbf e_{k}^{m})} = 
\begin{cases}  
3 & \textup{ if } m = 1\textup{ and } n = 2, \\
0 & \textup{ if } m = 3\textup{ and } n = 4, \\
0 & \textup{ if } m = 4\textup{ and } n = 3, \\
2 & \textup{ if } m = 3\textup{ and } n = 3, \\
2 & \textup{ if } m = 4\textup{ and } n = 4, \\
1 & \textup{ otherwise.}
\end{cases}\label{eq:pekrelation2}
\end{equation}
\normalsize
It follows from \eqref{eq:pekrelation1} and \eqref{eq:pekrelation2} that the identity \eqref{eq:ekrelation} can be found to hold after expanding the product \eqref{eq:factoredformaltrace} provided there holds the smaller set of relations
\begin{align}\label{eq:reducedrelations}
\begin{cases}
\mathbf e_{k-1}^m \mathbf e_{k}^n = 3 & \textup{if } m = 2 \textup{ and } n=1, \\
\mathbf e_k^m = 1 & \textup{otherwise.}
\end{cases}
\end{align}
This yields the expression
\small
\begin{align}\label{eq:traceraw}
\begin{split}
&\tr(E\rho^{(\vnu)}) = p^{(\vnu)} \\
&\quad\times \mathcal L\prod_{k=1}^{N-1} \big[P_{\vnu}(\mathbf e_{k}^{1})\bsigma_{2k-1,2k} + P_{\vnu}(\mathbf e_{k}^{2})\bsigma_{2k+1,2k+2}+ 2P_{\vnu}(\mathbf e_{k}^{3})\big]
\end{split}
\end{align}
\normalsize
where we have introduced vector quantities $\bsigma_{ij}$ and perform the multiplication in the commutative algebra $\mathcal A$ defined by $\bsigma_{ij}^2 = 3$.  The map $\mathcal L:\mathcal A\to\mathbb R$ is the linear map defined by $\bsigma_{ij}\mapsto 1$.  Specifically, we have put $\mathbf e_{k}^{1} = \bsigma_{2k-1,2k}$ and $\mathbf e_{k}^{2}=\bsigma_{2k+1,2k+2}$, and note that if substitute $P_\vnu(\mathbf e_k^4) = P_\vnu(\mathbf e_k^3)$ and put $\mathbf e_{k}^{3}= \mathbf e_{k}^4=1$ in \eqref{eq:factoredformaltrace} then upon expanding the product we obtain the required relations \eqref{eq:reducedrelations} provided $\bsigma_{ij}^2 = 3$ to account for all products $\mathbf e_{k-1}^{2}\mathbf e_{k}^{1}$ and otherwise we substitute $\bsigma_{ij}=1$.

Finally, we use the reduced state in Section \ref{sec:reducedstate} to calculate the required coefficients $P_\vnu(\mathbf e_k^m)$ yielding the result
\small
\begin{align}
\tr(E\rho^{(\vnu)}) = p^{(\vnu)}\mathcal L&\prod_{k=1}^{N-1} \betab_{2k,2k+1}^{(\nu_k,\nu_{k+1})}
\end{align}
\normalsize
where the non-vanishing factors $\betab_{2k,2k+1}^{(\nu_k,\nu_{k+1})}$ are given by \eqref{eq:errcoeffs1}-\eqref{eq:beta22} in the main text.  We thus obtain the efficiency of a $(2N-2)$-fold coincidence yielding a successful chain of BSMs as
\begin{equation}\label{eq:bsmeff}
\bar\eta_{1,2N} = \tr(E\hat\rho) = \sum_{\vnu} \tr(E\rho^{(\vnu)}) = \sum_{\vnu} p^{(\vnu)}\beta^{(\vnu)}
\end{equation}
where $\beta^{(\vnu)}$ is given by 
\begin{equation}\label{eq:errcoeffapp}
\cbeta^{(\vnu)} = \mathcal L\prod_{k=1}^{N-1} \betab_{2k,2k+1}^{(\nu_{k},\nu_{k+1})}
\end{equation}
and the sum is taken over all source pair emission numbers $\vnu\in\{0,1,2\}^N$.  Omitting the sequences with $\vnu_1=\vnu_N=0$ yields the 2$N$-fold coincidence efficiency 
\begin{equation}
\bar\eta_{AB} = \sum_{\substack{\vnu \\ \nu_1,\nu_N > 0} } p^{(\vnu)}\beta^{(\vnu)}
\end{equation}
if one assumes a heralding signal is used at the receivers $A$ and $B$ to remove the vacuum component.
 
The entire derivation above holds in the same form for the ABSM protocol, provided one uses the input modes $c_i,d_i$ in channels $i=2k,2k+1$ for $k$ even.  Changing the basis of the corresponding modes in the reduced state obtained in Section \ref{sec:reducedstate}, the relations \eqref{eq:pekrelation1}-\eqref{eq:reducedrelations} are modified accordingly with the only difference found in that the relation $\bsigma_{ij}^2=3$ must be replaced with $\bsigma_{ij}^2=0$. 

\subsubsection{Bell state fidelity}

In order to calculate the Bell state fidelity, we consider again the measurement POVM $\tilde E^\pm_k$ for each BSM.  Since the fidelity depends on the post-measurement state, some basic assumptions are required on the measurement operators $M_{m}^\pm$ which realize the POVM via the Kraus representation
\begin{equation}
\tilde E^- = \prod_{k=1}^{N-1} \tilde E^-_k = \sum_m M_m^\dagger M_{m}
\end{equation}
where we shall restrict our attention to the outcome consisting of $\tilde E_k^-$ in every BSM so that the state produced in the outer modes $a_1,b_1,a_{2N},b_{2N}$ is nominally the Bell state $|\Psi_{1,2N}^{-}\rangle\langle\Psi_{1,2N}^{-}|$.  Specifically, we require that the measurement operators $M_{m}$ commute with the Bell state projectors $|\Psi_{1,2N}^{\pm}\rangle\langle\Psi_{1,2N}^{\pm}|$ so that the fidelity of the state produced by the measurement realizing the POVM $\tilde E^-$ can be written
\begin{align}
F &= \frac{\sum_{m}\tr\big( M_{m}\hat\rho M_{m}^\dagger |\Psi_{1,2N}^{-}\rangle\langle\Psi_{1,2N}^{-}| \big)}{\sum_{m}\tr\big( M_{m}\hat\rho M_{m}^\dagger\big)} \\
&=\frac{\tr\big(\tilde E^-\hat\rho|\Psi_{1,2N}^{-}\rangle\langle\Psi_{1,2N}^{-}| \big)}{\tr\big(\tilde E^-\hat\rho\big)}
\label{eq:fidelitykraus}
\end{align}
independently of the specific operators $M_m$ which realize the POVM $\tilde E^-$.  This is a reasonable assumption on the operators $M_m$ by virtue of the fact that the BSM measurement apparatus should not directly interact with the optical modes $a_1,b_1,a_{2N},b_{2N}$.

The post-measurement normalization in \eqref{eq:fidelitykraus} was calculated in the previous section, and is given by 
\begin{equation}
\tr(\tilde E^- \hat\rho) = \frac{1}{2^{N-1}}\tr(\tilde E \hat\rho) = \frac{1}{2^{N-1}} \sum_{\vnu}p^{(\nu)}\beta^{(\vnu)}
\end{equation}
since here we count only the outcome $E^-$ which accounts for one of $2^{N-1}$ distinct successful BSM chain outcomes which all occur with equal probability (from symmetry considerations).  It remains to calculate the numerator
\begin{equation}\label{eq:etaabminus}
\eta_{AB}^- = \tr\big(\tilde E^-\hat\rho|\Psi_{1,2N}^{-}\rangle\langle\Psi_{1,2N}^{-}| \big).
\end{equation}

To this end, we can again expand the density matrix using \eqref{eq:dmdecompnu}
\small
\begin{equation}
\tr\big(\tilde E^-\hat\rho|\Psi_{1,2N}^{-}\rangle\langle\Psi_{1,2N}^{-}| \big) = \sum_{\vnu} \tr\big(\tilde E^-\hat\rho^{(\vnu)}|\Psi_{1,2N}^{-}\rangle\langle\Psi_{1,2N}^{-}| \big).
\end{equation}
\normalsize
We now make several observations.  First, we note that due to the Bell state projector the trace vanishes for all $\vnu$ except those with $\nu_1=\nu_N=1$, and so we henceforth consider only vectors $\vnu$ satisfying this condition.  Second, we note that for all such $\vnu$ with at least one 0-pair emission $\nu_i=0$, the trace satisfies 
\small
\begin{equation}
\tr\big(\tilde E^-\hat\rho^{(\vnu)}|\Psi_{1,2N}^{-}\rangle\langle\Psi_{1,2N}^{-}| \big) = \frac{1}{4}\tr\big(\tilde E^-\hat\rho^{(\vnu)} \big).
\end{equation}
\normalsize
This follows from general considerations, since there are no correlations between the optical modes $a_1,b_1$ and $a_{2N},b_{2N}$ if there is no physical interaction between two of the neighboring sources, leaving these modes in a completely mixed 2-photon state.  More generally, since the loss modes record when neighboring sources do not interact, the above relation holds for any tensor product of reduced states $\rho_{m;n}^{(\nu)}$ given in Section \ref{sec:reducedstate} such that at least one of the factors has $m=0$ or $n=0$.  It thus remains only to consider states of the form
\begin{equation}
\rho^{(\vnu)}_{1;1} = \rho_{1;1}^{(\nu_1)}\otimes \rho_{1;1}^{(\nu_2)}\otimes ... \otimes \rho^{(\nu_N)}_{1;1}
\end{equation}
with $\nu_i\geq 1$ and $\nu_1=\nu_N=1$.

Substituting \eqref{eq:rho111} and \eqref{eq:rho112}, we obtain an expression of the form
\small
\begin{align}
\begin{split}
&\tr\big(\tilde E^-\rho^{(\vnu)}|\Psi_{1,2N}^{-}\rangle\langle\Psi_{1,2N}^{-}| \big) \\
&= p^{(\vnu)}\prod_{i = 1}^{2N}\eta_i\prod_{\nu_k=2}4(1-\eta_{2k-1})(1-\eta_{2k}) \\
&\times\Big( \Big[ \frac{2}{3}\Big]^n \tr\big(\tilde E^-\hat\rho^-\big)  + \Big(1-\Big[\frac{2}{3}\Big]^n\Big)\tr(\tilde E^-\hat\rho^{\textup{mix}} |\Psi_{1,2N}^{-}\rangle\langle\Psi_{1,2N}^{-}|)\Big)
\end{split}
\end{align}
\normalsize
where $\hat\rho^-$ is the density matrix corresponding to the pure state $|\Psi_{12}^-\rangle|\Psi_{34}^-\rangle\cdots |\Psi_{2N-1,2N}^-\rangle$,  the remainder $\hat\rho^{\textup{mix}}$ is a normalized state consisting of a sum of tensor products of states $|\Psi_{ij}^-\rangle\langle \Psi_{ij}^-|$ with the state $\hat \rho_{1;1}^{\textup{mix}}$ in \eqref{eq:rho112}, and $n$ is the number of double-pair emissions $\nu_k=2$ in $\vnu$.  Similarly to the states with a zero-pair emission, any tensor product with the state $\hat\rho^{\textup{mix}}_{1;1}$ produces a completely mixed 2-photon state with one photon in each of the mode-pairs $a_1,b_1$ and $a_{2N},b_{2N}$, and so we again have
\begin{equation}\label{eq:rhomixfid}
\tr(\tilde E^-\hat\rho^{\textup{mix}} |\Psi_{1,2N}^{-}\rangle\langle\Psi_{1,2N}^{-}|\Big) = \frac{1}{4}\tr(\tilde E^-\hat\rho^{\textup{mix}}\big).
\end{equation}
Combining \eqref{eq:etaabminus}-\eqref{eq:rhomixfid}, the result is
\begin{equation}
\eta_{AB}^{-}=\frac{1}{2^{N-1}}\sum_{\substack{\vnu\in\{0,1,2\}^N \\ \nu_1=\nu_N=1}} p^{(\vnu)}\Big(\frac{1}{4} \cbeta^{(\vnu)} + \frac{3}{4}\Big[\frac{2}{3}\Big]^n \hat\beta^{(\vnu)}\Big) 
\end{equation}
where $\beta^{(\vnu)}$ is given by \eqref{eq:errcoeffapp} and counts the probability of a successful BSM given $\vnu$ photon pairs emitted by each source, and $\hat\beta^{(\vnu)}$ counts only the contribution from tensor products $\hat\rho^{(\vnu)}_{1;1}$, which can be obtained following the derivation of the previous section by setting $\bsigma_{ij}=0$ in \eqref{eq:traceraw}.  In the case that no heralding signal is employed by the receivers this yields the fidelity 
\begin{equation}
F = \frac{\eta_{AB}}{\bar\eta_{1,2N}}
\end{equation}
where $\eta_{AB}=2^{N-1}\eta_{AB}^-$.  The case of receivers capable of vacuum filtering can be treated simply by removing the vacuum terms from the post-measurement normalization (since they are already projected out of $\eta_{AB}^-$) to obtain the vacuum-filtered fidelity
\begin{equation}
F = \frac{\eta_{AB}}{\bar\eta_{AB}}.
\end{equation}

\subsection{Calculation of the optimal pair emission probabilities for a Type-II elementary link}

The optimal pair emission probabilities---used to determine the maximum efficiency of an imbalanced Type-II elementary link with probabilistic sources in Section \ref{sec:absmgain}---were calculated using the method of Lagrange as discussed in the main text.  As there are several subtleties to the calculation, we present the full derivation here.

Recall that the objective is to maximize the function
\begin{equation}
\hat\eta_{AB}(\Delta f) = \max_{p_{12},p_{34},p_{56}}\Big\{\eta_{AB} : 
\begin{array}{ll} F \geq 1-\Delta f\\
 p_{ij}\leq 8/27\end{array} \Big\}
\end{equation}
where $\eta_{AB}$ is given by \eqref{eq:concatefftrue} with $N=3$ and $F = \eta_{AB}/\bar\eta_{AB}$.  For fixed channel losses, the functions $\eta_{AB}$ and $F$ are parameterized by three parameters $(p_1,p_2,p_3)=(p_{12},p_{34},p_{56})$ representing the single-pair emission probability, provided we assume the multi-pair emissions are related to the single-pair emissions by \eqref{eq:npaird}.  To facilitate the calculation, we approximate this relationship by
\begin{align}
p_{ij}^{(2)} &\simeq \gamma p_{ij}^2 \\
p_{ij}^{(0)} &\simeq 1 - p_{ij} - \gamma p_{ij}^2
\end{align}
with $\gamma = 3/4$, which holds for $p_{ij}\ll 1$.  The second approximation that we make is that rather than maximizing $\eta_{AB}$ directly, we instead maximize the product $\tilde p = p_1p_2p_3$ which is roughly proportional to $\eta_{AB}$ for $\Delta f \ll 1$.

The latter approximation allows us to use the Lagrangian constraint $\partial_i F = \lambda\partial_i \tilde p$ (where $\partial_i = \partial/\partial p_{i}$) in the form
\begin{equation}\label{eq:lconstraint}
p_i\partial_i F = \lambda p_i \partial_i \tilde p = \lambda \tilde p
\end{equation}
where the right hand side is independent of $i$ yielding
\begin{equation}\label{eq:constraint}
p_1\partial_1 F = p_2\partial_2 F = p_3\partial_3 F.
\end{equation}
Direct calculation reveals that
\begin{align}
p_i\partial_i \eta_{AB} &= \eta_{AB} + \epsilon_i \\
p_i\partial_i \bar\eta_{AB} &= \bar\eta_{AB} + \bar\epsilon_i
\end{align}
where the remainders $\epsilon_i,\bar\epsilon_i$ are linear combinations of the error terms $p^{(\vnu)}\beta^{(\vnu)}$.  If we restrict to terms of order at most 4 in the parameters $p_i$ including terms $\beta^{(\vnu)}$ consisting of at most 4 photon pair emissions from the 3 sources---the remainders $\epsilon_i,\bar\epsilon_i$ are given by
\begin{align}
\epsilon_1 &= \epsilon_3 = 0, \\
\epsilon_2 &= \frac{1}{4}\gamma p_1p_2^2p_3 \beta^{(1,2,1)} + \frac{1}{2}\gamma p_1p_2^2p_3 \hat\beta^{(1,2,1)}.
\end{align}
and 
\begin{align}
\bar\epsilon_1 &= \gamma p_1^2p_2p_3 \beta^{(2,1,1)} + \gamma^2 p_1^2p_3^2\beta^{(2,0,2)}, \\
\bar\epsilon_2 &= \gamma p_1p_2^2p_3 \beta^{(1,2,1)} - \gamma^2 p_1^2p_3^2\beta^{(2,0,2)}, \\
\bar\epsilon_3 &= \gamma p_1p_2p_3^2 \beta^{(1,1,2)} + \gamma^2 p_1^2p_3^2\beta^{(2,0,2)}.
\end{align}
We thus obtain
\begin{equation}\label{eq:fidelityderivative}
p_i \partial_i F = \frac{\epsilon_i - \bar\epsilon_iF}{\bar\eta_{AB}}.
\end{equation}
Combined with \eqref{eq:constraint} and substituting $F\simeq 1$ (using $\Delta f\ll 1$) we obtain a set of equations
\begin{equation}\label{eq:prequadraticequations}
p_1b_{12} = p_2 b_{34} - \frac{p_1p_3}{p_2} b_2 = p_3 b_{56} 
\end{equation}
where
\small
\begin{align}
b_{12} &= \beta^{(2,1,1)}, \\
b_{56} &= \beta^{(1,1,2)}, \\
b_{34} &= \frac{3}{4}\beta^{(1,2,1)}-\frac{1}{2}\hat\beta^{(1,2,1)}, \\
b_{2} &= 2\gamma \beta^{(2,0,2)}.
\end{align}
\normalsize
The first equation in \eqref{eq:prequadraticequations} can be simplified by substituting $p_{3}=p_1b_{12}/b_{56}$ and solving the resulting quadratic in $x=(p_1/p_2)$ to obtain
\begin{align}
\begin{split}
p_{1}b_{12} = p_{2}b_{34}w(b) = p_{3}b_{56} \equiv p
\end{split}
\end{align}
where $w(b)=2/(1+\sqrt{1+8\gamma b})$ with
\begin{align}\label{eq:bparameter}
\begin{split}
b &= \frac{1}{2\gamma}\frac{b_{34}b_2}{b_{12}b_{56}}.
\end{split}
\end{align}

The absolute pair probabilities for fixed $F=1-\Delta f$ are then determined by the parameter $p$ which can be obtained by inverting $F(p)$.  Noting that $F$ has a non-vanishing derivative with respect to $p$ we can use the linear approximation
\begin{equation}
p = -\Big(\frac{\partial F}{\partial p}\Big\vert_{p=0}\Big)^{-1} (\Delta f) + O\big( (\Delta f)^2 \big).
\end{equation}
The derivative is obtained from $\partial_i F$---calculated in \eqref{eq:fidelityderivative}---via the chain rule.  Evaluating at $p=0$ yields
\begin{equation}
\frac{\partial F}{\partial p}\Big\vert_{p=0} = \frac{3\gamma}{\beta^{(1,1,1)}}\big(1 + \gamma b w(b)\big)
\end{equation}
using the limit $\bar\eta_{AB}\sim p_1p_2p_3 \beta^{(1,1,1)}$ as $p\to 0$.

The optimization of the Type-II link with cascaded sources discussed in Section \ref{sec:absmgain} is modeled as a concatenated entanglement swap with $N=5$ sources parameterized by 3 parameters $p_1=p_{12}=p_{34}$, $p_2=p_{56}$ and $p_3 =p_{78}=p_{9,10}$.  In this case, we optimize the function $\tilde p = p_1^2p_2p_3^2$ which modifies the Lagrangian constraint \eqref{eq:lconstraint} to the form
\begin{equation}
\frac{1}{m_i} p_i\partial_i F = \lambda \tilde p
\end{equation}
where $m_i$ is the multiplicity given by $m_2=1$ and $m_1 = m_3 = 2.$  The rest of the calculation follows in a similar fashion with the results given in Section \ref{sec:absmgain}.
\vfill

\end{document}